# Short-term Variability of the Sun –Earth System: An Overview of Progress Made during the CAWSES-II Period

Nat GOPALSWAMY[1]

Corresponding author

Email: nat.gopalswamy@nasa.gov

Bruce TSURUTANI[2]

Email: bruce.t.tsurutani@jpl.nasa.gov

Yihua YAN[3]

Email: yyh@nao.cas.cn

[1] Solar Physics Laboratory, Code 671, Heliophysics Division, NASA Goddard Space Flight Center, Greenbelt, MD 20771, USA

[2] Jet Propulsion Laboratory, California Institute of Technology, Pasadena, California USA

[3] Key Laboratory of Solar Activity, National Astronomical Observatories, Chinese Academy of Sciences, Beijing 100012, China

Science section: 1) space and planetary sciences

## Abstract

This paper presents an overview of results obtained during the CAWSES-II period on the short-term variability of the Sun and how it affects the near-Earth space environment. CAWSES-II was planned to examine the behavior of the solar-terrestrial system as the solar activity climbed to its maximum phase in solar cycle 24. After a deep minimum following cycle 23, the Sun climbed to a very weak maximum in terms of the sunspot number in cycle 24 (MiniMax24), so many of the results presented here refer to this weak activity in comparison with cycle 23. The short-term variability that has immediate consequence to Earth and geospace manifests as solar eruptions from closed-field regions and high-speed streams from coronal holes. Both electromagnetic (flares) and mass emissions (coronal mass ejections – CMEs) are involved in solar eruptions, while coronal holes result in high-speed streams that collide with slow wind forming the so-called corotating interaction regions (CIRs). Fast CMEs affect Earth via leading shocks accelerating energetic particles and creating large geomagnetic storms. CIRs and their trailing high speed streams (HSSs), on the other hand, are responsible for recurrent small geomagnetic storms and extended (days) of auroral zone activity, respectively. The latter lead to the acceleration of relativistic magnetospheric "killer" electrons. One of the major consequences of the weak solar activity is the altered physical state of the heliosphere that has serious implications for the shock-driving and storm-causing properties of CMEs. Finally, a discussion is presented on extreme space weather events prompted by the 2012 July 23 super storm event that occurred on the backside of the Sun. Many of these studies were enabled by the simultaneous availability of remote-sensing and in-situ observations from multiple vantage points with respect to the Sun-Earth line.

# 1. Introduction

The second phase of the Climate and Weather of the Sun-Earth System (CAWSES-II) was organized into task groups (TGs). Task Group 3 (TG3) was focused on the short-term variability of the Sun-Earth system. Solar variability on timescales up to 11 years was relevant to TG3. The relevant variability occurs in the mass and electromagnetic outputs of the Sun. The mass output has three forms: the solar wind, coronal mass ejections (CMEs), and solar energetic particles (SEPs). The electromagnetic output consists of the quasi-steady black body radiation with the superposition of flare emission. The mass and electromagnetic emissions are often coupled: flares and CMEs represent two different manifestations of the energy release from solar source regions (e.g., Asai et al. 2013). SEPs are accelerated in fast CME-driven shocks as well as by flare reconnection (see e.g., Reames 1999; 2013). The source regions of flares and CMEs are closed magnetic field regions such as active regions and filaments (see e.g. Srivastava et al. 2014). Active regions consist of sunspots of opposite polarity at the photospheric level. Filament regions do not have sunspots, but consist of opposite polarity magnetic patches. The main difference between the two regions is the magnetic field strength: hundreds of gauss in the sunspot regions vs. tens or less gauss in filament regions. Sunspots also contribute to the variability in total solar irradiance (TSI). While sunspots decrease the TSI, the plages that surround the sunspots increase it, resulting in higher TSI when the Sun is more active.

The solar wind has fast and slow components. The fast component is of particular interest because it can compress the upstream slow solar wind forming a corotating interaction region (CIR). Coronal holes, the source of the fast solar wind, also exhibit remarkable variability in terms of their location on the Sun and size. Perhaps even more important than the CIR is the high speed stream proper. It carries large nonlinear Alfvén waves, whose southward components cause reconnection at the magnetopause resulting in continuous sporadic plasmasheet injections into the nightside magnetosphere. These injections of anisotropic ~10 to 100 keV electrons cause the growth of an electromagnetic wave called "chorus" and the chorus interacts with the ~100 keV electrons accelerating them to MeV energies (Tsurutani et al., 2006, 2010; Thorne et al., 2013).

CMEs are launched into the solar wind, so the two mass outputs interact and exchange momentum affecting the propagation characteristics of CMEs in the interplanetary medium. CMEs also interact with the upstream heliospheric current sheet and other material left over from other injections. The variability manifested as solar flares, CMEs, SEPs, and high-speed solar wind streams directly affects space weather on short time scales. As noted above, all these phenomena are coupled not only near the Sun, but throughout the inner heliosphere, including geospace and Earth's ionosphere and atmosphere where the impact can be felt (Verkhoglyadova et al., 2014; Mannucci et al. 2014; Tsurutani et al., 2014).

## 2. Methods

This review covers the second phase of the CAWSES program, known as CAWSESS II, which began in 2009 and ended in 2013, roughly covering the rise to the maximum phase of solar cycle 24. There is strong evidence showing that solar cycle 24 is a relatively weak cycle (Tan 2011; Basu 2013). The birth of solar cycle 24 was remarkable in that the Sun emerged from an extremely deep minimum. The maximum phase of cycle 24 is of particular interest because of the sunspot number was rather small (roughly half of the cycle-23 peak). The weak solar cycle resulted in a milder space weather, but there were other complications such as longer-living space debris due to the reduced atmospheric drag. SCOSTEP conducted a year-long campaign known as "MiniMax24" to document solar events and their geospace impact during the mild maximum phase of cycle 24. Additionally, the solar mid-term and long-term quasi-periodic cycles and their possible relationships with planetary motions from long-term observations of the relative sunspot number and microwave emission at frequency of 2.80 GHz were also investigated, and it was suggested that the mid-term solar cycles (periods < 12 yr) are closely related to the motions of the inner planets and of Jupiter (Tan and Cheng 2013).

This paper highlights some key results obtained on the variable phenomena in the Sun-Earth system during the CAWSES-II time frame. Detailed treatment of topics relevant to CAWSES-II can be found in the Living Reviews in Solar Physics: Hathaway (2010), Charbonneau (2010), Petrovay (2010), Chen (2011), Shibata and Magara (2011), Aschwanden (2011), Webb and Howard (2012), Usoskin (2013), Potgieter (2013), Lockwood (2013), Owens and Forsyth (2013).

## 3. Review

### 3.1 The Weak Solar Cycle 24 and Its Consequences

CAWSES-II focused on solar variability as the Sun approached the maximum of cycle 24. The rise phase of cycle 24 was already interesting because the Sun emerged from a deep solar minimum that gained particular interest among solar-terrestrial scientists (Selhorst et al. 2011;

Tsurutani et al., 2011a; Dasso et al. 2012; Gopalswamy et al. 2012a; Solomon et al. 2013; Lean et al. 2014; Potgieter et al. 2014). Solar activity is typically represented by the international sunspot number (SSN), but there are many other measures, which are needed for a complete understanding of the solar variability. In particular, measurements of the polar field strength, tilt angle of the heliospheric current sheet (Smith et al., 1978), latitudes of filament locations, and coronal streamers all provide complementary information on the solar activity as the Sun climbed towards its weak maximum around 2013.

**3.1.1 Solar Polar Field Strength**

Figure 1 shows the equatorial (sunspot number – SSN, active region microwave brightness temperature - Tb) and polar (magnetic field strength) manifestations of solar cycle 24, in comparison with cycle 23. The sunspot number (SSN) and the equatorial microwave Tb indicate the strength of the activity cycle. Clearly the decline in SSN from cycle 23 to 24 is very pronounced. The Tb decline is also evident. The Tb from the northern and southern active region belt shows marked asymmetry around the solar maxima. The northern AR emission shows only a single peak during the first SSN peak, while the southern AR Tb shows double peaks, the first peak being weaker than the second peak in the two cycles. The combination of northern and southern peaks in SSN gives rise to the well-known Gnevishev gap. The Gnevishev gap was somewhat wider during the cycle 24 maximum. The polar field strength during the prolonged cycle 23/24 minimum was considerably smaller than that during the cycle 22/23 minimum. During solar minima, the polar field strength (B) reaches its peak values and vanishes at maxima, changing sign at the end of maxima (Selhorst et al. 2011; Tsurutani et al., 2011a; Gopalswamy et al. 2012a; Shimojo 2013; Nitta et al. 2014; Mordvinov and Yazev 2014). The maximum phase is indicated by the vanishing polar field strength. The polar B in Fig. 1 thus suggests that the maximum phase of cycle 24 is almost over. Note that the arrival of the maximum phase is not synchronous in the northern and southern hemispheres for cycles 23 and 24 (see also Svalgaard and Kamide 2013). The lag in the southern hemisphere is more pronounced in cycle 24. Polar microwave Tb also declined significantly between the cycle 22/23 and 23/24 minima (Gopalswamy et al. 2012a). During solar maxima, the polar Tb drops to the quiet Sun values (~$10^4$ K) because the polar coronal holes disappear. It must be pointed out that the southern polar field was stronger during both 22/23 and 23/24 minima and accordingly the active region Tb was higher in the southern hemisphere during the cycle 23 and cycle 24 maxima, indicating a close relationship between the polar field strength during a minimum and the activity strength in the following maximum.

### 3.1.2 Implications for the Solar Dynamo

According to the Babcock-Leighton mechanism of the solar dynamo, the polar field strength of one cycle determines the strength of the next cycle. The so-called polar precursor method of predicting the strength of a solar cycle using the peak polar field strength of the preceding minimum has been fairly accurate (see e.g., Svalgaard et al. 2005; Jiang et al. 2013a; Muñoz-Jaramillo et al. 2013; Zolotova and Ponyavin 2013). Recent discussion on the precursor method can be found in Pertrovay (2010) and Pesnell (2014) among others. In addition to the traditional polar field measurements, proxies such as H-alpha synoptic charts (Obridko and Shelting 2008) and the polar microwave Tb (Gopalswamy et al. 2012a) can also be used to predict the strength of the activity cycle. The polar microwave Tb is exceptionally good because it is highly correlated (correlation coefficient $r = 0.86$) with the polar field strength: $B = 0.0067Tb - 70$ G (Gopalswamy et al. 2012a).

The enhanced low-latitude Tb between 1997 and 2008 corresponds to the solar activity in cycle 23 representing the toroidal field (see also Selhorst et al. 2014). The enhanced high-latitude Tb between 1992 and 1996 represents the poloidal field. The correlation between the high- and low-latitude microwave Tb, averaged over Carrington rotation periods is shown in Fig. 2 using data from cycle 22/23 minimum and cycle 23 maximum. Clearly, the correlation is rather high ($r=0.74$ for the northern hemisphere and 0.82 for the southern hemisphere). However, the correlation plots look very different in the northern and southern hemispheres. The maximum correlation occurs for a lag of 75 rotations in the northern hemisphere (~5.7 years) and 95 in the south (7.2 years in the south). The north-south asymmetry noted before for the arrival of maximum phase is also clear in the correlation plots. Thus, the Nobeyama observations provide a strong observational support to the idea that the poloidal field of one cycle decides the strength of the next cycle (toroidal or sunspot field). Furthermore, the Nobeyama data provides a more detailed time structure (Carrington rotation) compared to those (a solar cycle) used in other studies. This finding for cycle 23 can be tested for cycle 24 when it ends in the next few years.

How do we understand the weak cycle 24? Jiang et al. (2013b) considered several possibilities such as (i) the accuracy of SSN, (ii) sunspot tilt angle variation, and (iii) the variation in the meridional circulation during cycle 23. They were able to reproduce the lower polar field during the cycle 23/24 minimum using a 55% increase of the meridional flow in their model. They also found that a 28% decrease of the mean tilt angle of sunspot groups can explain the low polar field, but this would not be consistent with the observed time of polar field reversals. They concluded that the nonlinearities in the polar field source parameters and in the transport parameters play important roles in the modulation of the polar field.

### 3.1.3 Implications for the Long-term Behavior of the Sun

The Sun is known to have variability on time scales up to millennia (See Usoskin 2013 for a review). One obvious question is whether the weakening of the activity observed in cycle 24 will continue further. Javaraiah (2015) examined the north-south asymmetry of sunspot areas binned into 10º latitudes and examined various periodicities. They found periodicities of 12 and 9 years, respectively during low-activity (1890–1939) and high-activity (1940–1980) intervals. They also inferred that cycle 25 may be weaker than cycle 24 by ~31%. Several authors have discussed the possibility of a global minimum over the next several cycles (se e.g., Russell et al. 2013a; Lockwood et al. 2011; Steinhilber and Beer 2013; Zolotova and Ponyavin 2014; Ruzmaikin and Feynman 2014). Zolotova and Ponyavin (2014) reported that the protracted cycle 23 is similar to the cycles immediately preceding the Dalton and Gleissberg-Gnevyshev minima, suggesting that the Sun is heading towards such a grand minimum.

But the most important is that the diminished solar activity has immediate consequences for the society. When the Maunder Minimum occurred in the late 1600s, the technology was not seriously affected by the Sun. Today's technology is extensively coupled to solar activity, so the effect is readily recognized. For example, the weak solar activity has resulted in reduced atmospheric drag on satellites increasing their life time. On the other hand, space debris do not burn up quickly, thus posing additional danger to the operating satellites. The geomagnetic disturbances have been extremely mild, with the weakest level of geomagnetic storms since the space age.

### 3.1.4 The Weakest Geomagnetic Activity on Record: Cycle 23 Minimum

Figure 3 shows from top to bottom: the sunspot number (Rz), the 1 AU interplanetary magnetic field magnitude (Bo), the Oulu Finland cosmic ray count rate (the local vertical geomagnetic cutoff rigidity is ~0.8 GV), the solar wind speed (Vsw) and the ap geomagnetic index. The vertical dashed green lines give the official dates of the solar minima between cycle 22 and 23 and cycle 23 and 24 (Hathaway, 2010). The vertical blue lines give the geomagnetic ap index minima. The horizontal red lines have been added to the figure to guide the reader. From top to bottom, the lines are the zero value for Rz, 5 nT for Bo, 6500 cts/min for the cosmic ray flux, 400 km/s for Vsw and 10 nT for ap. We call the reader's attention to the long delay between the sunspot minima and the geomagnetic activity minima. This occurs in both solar cycle minima. The 2010 geomagnetic ap minimum is the lowest since the index began to be recorded.

The figure shows that cycle 23 extended from 1996 to 2008 and is the longest in the space era (12.6 years). For comparison, the length of solar cycles 20 through 22 were 11.7, 10.3 and 9.7 years, respectively. The values below the red lines have been shaded for emphases (in the case of cosmic rays, the values above the red line are shaded). It can be noted that the Bo, Vsw and ap index values for the cycle 23 minimum are considerably lower than the cycle 22 minimum values. The minimum in ap is broad and extends from day 97, 2008 until day 95, 2010. The onset and end times are somewhat arbitrary. There is a minimum geomagnetic activity interval in cycle 22 (day 106, 1996 to day 23, 1998).

Figure 4 shows the solar wind, the IMF magnetic field magnitude, the interplanetary epsilon parameter (Perrault and Akasofu, 1978) and the ap index from 2008 through the first part of 2010. This interval is noted for a general lack of CMEs (and magnetic storms) and the dominance of high speed streams (top panel). What is unusual about this is the general decline in the peak solar wind speed starting in 2008 and extending to 2010. The peak solar wind speeds of high speed streams are typically 750 to 800 km/s at 1 AU and beyond (Tsurutani et al. 1995; Tsurutani and Ho, 1999), but here none of these streams have these magnitudes.

What is the cause of this extremely low geomagnetic activity between cycle 23 and cycle 24? It was found that coronal holes during this phase of the solar cycle are small and located near middle latitudes (De Toma, 2011). This caused the solar wind speed from coronal holes to be weak and the magnetic field variances to be particularly low (not shown). A schematic to indicate all of these features are shown in Fig. 5.

It is surmised that nothing has changed on the speed of the high speed streams emanating from coronal holes during solar minimum. The terminal speed is still ~750 to 800 km/s. However this is the speed for the central portion of the hole. As the high speed stream expands into interplanetary space, it does not simply propagate radially outward, but expands into nearby space, leading to "super-radial" expansion as shown in the schematic of Fig. 5. At the sides of the high speed stream, the speed and the amplitude of the entrained Alfvén waves will be reduced. This is the portion of the high speed streams that hit the Earth's magnetosphere. Thus the low solar magnetic fields, the lack of CMEs, the midlatitude location of small coronal holes all contribute to the all-time minimum in the geomagnetic activity between 2008 and 2009. It is noted that in Fig. 3, a similar feature can be noted in the cycle 22 minimum, but the feature is less prominent.

## 3. 2 Coronal Mass Ejections and Flares

### 3.2.1 Origin of Solar Eruptions

Although it is well established that CMEs and their interplanetary manifestations, ICMEs, and flares originate from closed-field regions on the Sun such as active regions and filament regions, the current level of understanding is not sufficient to predict when an eruption might occur in such a region. Two basic processes seem to be involved: energy storage and triggering. The energy storage can be identified from non-potentiality of the source region such as magnetic shear or accumulated helicity (Tsurutani et al., 2009; Kazachenko et al. 2012). Zhang et al (2012) studied the magnetic helicity of axisymmetric power-law force-free fields and focused on a family whose surface flux distributions are defined by self-similar force-free fields. The results suggest that there may be an absolute upper bound on the total magnetic helicity of all bipolar axisymmetric force-free fields.

In addition to the energy storage, a trigger in the form of a magnetic disturbance seem to be necessary, which causes a pre-eruption reconnection (Kusano et al. 2012). These authors suggest that observing these triggers is important for predicting eruptions and that we can predict eruptions only by a few hours before the eruption. For longer term predictions, one has to resort to probabilistic methods. Huang et al (2011) presented a study of a coronal mass ejection (CME) with high temporal cadence observations in radio and extreme-ultraviolet (EUV). The radio observations combined imaging of the low corona with radio spectra in the outer corona and interplanetary space. They found that the CME initiation phase was characterized by emissions that were signatures of the reconnection of the outer part of the erupting configuration with surrounding magnetic fields. Later on, a main source of emission was located in the core of the active region, which is an indirect signature of the magnetic reconnection occurring behind the erupting flux rope. Energetic particles were also injected in the flux rope and the corresponding radio sources were detected. Other radio sources, located in front of the EUV bright front, traced the interaction of the flux rope with the surrounding fields. They found that imaging radio emissions in the metric range can trace the extent and orientation of the flux rope which was later detected in interplanetary space.

**3.2.2 Long-term Behavior of CME Rates**

Although CMEs were discovered in 1971 (Tousey 1973), understanding their long-term behavior became possible only after the launch of the Solar and Heliospheric Observatory (SOHO). The continuous observations from the Large Angle and Spectrometric Coronagraph (LASCO) on board SOHO since 1996 constitute a uniform and extended data set on CMEs. Figure 6 shows a plot of the CME rate and speed averaged over Carrington rotation periods (27 days) for cycle 23 and 24, including the prolonged minimum between the two cycles. Only CMEs of width 30°or larger have been included in the plots; including narrower CMEs would increase the rate even higher in cycle 24. We used the width criterion to avoid variability due to manual identification by different people and the change in SOHO/LASCO image cadence in 2010. We see that the CME rate over the first five years in each cycle is not drastically different, even though the sunspot number dropped significantly. The prolonged cycle 23/24 minimum had low CME rate (but non-zero), similar to the cycle 22/23 minimum. The average CME speed decreased significantly during the 23/24 minimum compared to that during the 22/23 minimum. However, the average speeds during the cycle 23 and 24 maxima were not significantly different.

Figure 7 shows a detailed comparison between the corresponding epochs of cycles 23 and 24. The SSN averaged over the first five years in each cycle dropped from 68 to 38, which is a 44% reduction in cycle 24. On the other hand, the CME rate remained the same (2.09 in cycle 23 vs. 2.10 in cycle 24). This means the relation between SSN and CME rate changed in cycle 24 (the daily CME rate per SSN is greater in cycle 24), which will be discussed in section 3.2. There is ongoing debate to understand the reason for this difference: possible artifacts (Wang and Colaninno 2014; Lamy et al. 2014), changing strength of the poloidal field (Petrie 2012; 2013) or the altered state of the heliosphere (Gopalswamy et al. 2014a).

**3.2.3 Importance of CMEs for Space Weather**

For space weather effects, more energetic CMEs need to be examined. Gopalswamy et al. (2014b) started with flares of soft X-ray size ≥C3.0. This criterion avoids the effect of soft X-ray background level and its variability between the two cycles. For example ~20% of flares of size <C1 cannot be detected by GOES during cycle 23, while the corresponding fraction is 6% for cycle 24. However, for flares of size ≥C3.0, no flares go undetected. The source locations of flares were obtained from the online Solar Geophysical Data (SGD) or identified using near-surface observations from a number of sources: EUV images from SOHO, STEREO (Solar Terrestrial Relations Observatory), and SDO (Solar Dynamics Observatory), soft X-ray images from Yohkoh, microwave images from the Nobeyama Radioheliograph, and H-alpha images from various observatories. For each of these flares, the association of a CME was checked using the SOHO/LASCO catalog (http://cdaw.gsfc.nasa.gov, Gopalswamy et al. 2009a) to compile the properties of the associated CMEs. Only flares that occurred within 30º from the limb were considered, so that the speed and width measurements of CMEs were subject to minimal projection effects.

Figure 8 shows histograms of soft X-ray flares binned into various classes. Clearly, the number of flares with CME association depends on the flare size as is well known (Yashiro et al. 2005). The mean and median flare sizes are roughly the same in both cycles and the shape of the distributions are quite similar. The total number of flares of size ≥C3.0 was 664 and 554 for cycle 23 and 24, respectively over the first 62 months in each cycle. This corresponds to a reduction of ~17% in cycle 24. This reduction is not as large as the drop in SSN. The number of flares with CMEs was 273 for cycle 23 compared to 214 in cycle 24, which corresponds to a reduction of ~ 22% from cycle 23 to 24. Accounting for the lack of CME data for about 4 months in cycle 23 when SOHO was temporarily disabled, the reduction was 27%. Again the decline in the CME rate was not as drastic as the SSN. However, this is different from the same average CME rate found for cycles 23 and 24 (Fig. 7) suggesting that the reduction was in the number of energetic eruptions.

### 3.2.4 CME Speed and Width Distributions

The properties of the CMEs in the two cycles are compared in Figure 9. The speed distributions in the two cycles were similar with almost the same average (633 km/s vs. 614 km/s) and median (514 km/s vs. 495 km/s) speeds. On the other hand, the width distributions were significantly different: the average and median widths of non-halo CMEs in cycle 24 were much higher than those in cycle 23 over the corresponding epoch. The fraction of halo CMEs in cycle 23 was 3%, which is typical of the general population of CMEs (Gopalswamy et al 2010a). However, the halo fraction was 9% in cycle 24, three times larger than that in cycle 23.

The overabundance of cycle-24 halo CMEs was also observed in the general population. Halo CMEs are so-called because new material is observed all around the occulting disk in sky-plane projection (Howard et al 1982; Gopalswamy et al 2010a). Figure 10 shows the distribution of halo CMEs (binned over Carrington rotation periods) as a function of time. There were 199 halo CMEs (apparent width = 360º) during cycle 24 until the end of April 2014, amounting to ~3.06 CMEs per month. On the other hand, there were only 178 halos during the first 65 months of cycle 23, or 2.99 CMEs per month (adjusting for the 4 of months SOHO was not observing in cycle 23). Clearly, the halo CME occurrence rate in cycle 24 did not decrease at all (see Gopalswamy et al. 2015 for more details). For a given coronagraph, halo CMEs represent fast and wide (and hence energetic) CMEs (Gopalswamy et al. 2010b), further suggesting something peculiar about CME widths in cycle 24.

Table 1 compares the number of CMEs in cycles 23 and 24 under various categories (CME width, width and speed, and flare size). "All CMEs" includes every CME that was identified and measured. The largest difference was for the narrowest CMEs (width W<30º): the number of narrow CMEs in cycle 24 was a factor of 2 higher, which was the reason for the difference in the general population. These CMEs are generally not well defined and do not travel far from the Sun. There may be multiple reasons for the overabundance of narrow CMEs including measurement bias and the change in image cadence. The number of CMEs in cycles 23 and 24 became equal for wider CMEs. For CMEs with width ≥60º, the monthly rate was roughly the same in the two cycles. This was also true for halo CMEs. Fast and wide CMEs, on the other hand, showed a decline in number in cycle 24 as did CMEs associated with ≥C3.0 flares.

### 3.2.5 Anomalous Expansion of CMEs in Cycle 24
The speed vs. width scatterplot in Figure 11 further illustrates the different width distribution in cycle 24 (Gopalswamy et al. 2014a). The plot confirms the well-known linear relationship between speed and width of CMEs (faster CMEs are wider), but the slope is significantly larger in cycle 24. This means that the cycle-24 CMEs are significantly wider for a given speed. This also explains the high abundance of halo CMEs in cycle 24 even though there is a slight reduction in the number of energetic CMEs in this cycle.

### 3.2.6 CME Mass Distribution in Cycles 23 and 24
CME mass can be determined from LASCO images and is thought to be accurate within a factor of ~2. Since we are considering limb CMEs, the projection effects are minimal and the mass estimate is expected to be more accurate. The average masses shown in Fig. 12 is over the first 62 months of cycles 23 and are generally consistent with previous estimates (Gopalswamy et al 2010b; Vourlidas et al. 2011). However, the cycle-24 CME masses are smaller by a factor of ~3 (see Fig. 12). The cycle-24 value is also smaller than the mass averaged over the entire solar cycle 23 (Vourlidas et al. 2011). Gopalswamy et al. (2005a) used more than 4000 CMEs during 1996 to 2003 and found that the CME mass (M) and width (W) were correlated (r=0.63) with a regression equation: $\log M = 1.3 \log W + 12.6$. This relationship is also true for the limb CMEs used in Fig. 11: $\log M = 1.54 \log W + 12.4$ (cycle 23) and $\log M = 1.84 \log W + 11.5$ (cycle 24). The slope of the cycle-24 regression line is slightly larger. For a CME of ~60º width, the cycle-23 CME mass was larger by a factor of ~2.4. This result is consistent with the anomalous expansion of cycle-24 CMEs: a 60º wide CME in cycle 24 is equivalent to a narrower CME in cycle 23.

### 3.2.7 The Weak State of the Heliosphere

The inflated CME size in cycle 24 seems to be a direct consequence of the weak heliosphere, stemming from the weaker activity at the Sun. The physical parameters of the heliosphere all showed smaller values in the rise to the maximum phase of cycle 24. McComas et al. (2013) extended their earlier work (McComas et al. 2008) on the extended cycle 23/24 minimum to the rise phase of cycle 24 and shown that the solar wind densities, proton temperatures, dynamic pressures, and interplanetary magnetic field strengths were all diminished. Even the density fluctuations in the slow solar wind diminished significantly (Tokumaru et al. 2013). The effect was even felt at the heliospheric termination shock, whose size decreased by ~10 AU (J. Richardson, 2014, private communication). Gopalswamy et al. (2014a) reported that the total pressure (magnetic + kinetic) pressure, magnetic field, and the Alfven speed all declined significantly in cycle 24. Figure 13 compares several solar wind parameters measured at 1 AU and the same values extrapolated to the vicinity of the Sun. The reduced heliospheric pressure can readily explain the inflated CMEs in cycle 24. The drastic change in the state of the heliosphere between cycles 23 and 24 has important implications for space weather events (see later).

### 3.2.8 Forbush Decrease

Forbush decrease (FD) represents the reduction in the intensity of galactic cosmic rays (GCRs)

as detected by neutron monitors and muon detectors due to solar wind disturbances (see e.g., Munakata ET AL. 2005; Dumbovic et al. 2012; Arunbabu et al. 2013; Ahluwalia et al. 2014; Belov et al. 2014). Both CMEs and CIRs cause a FD, but the amplitude is significantly higher for CMEs than for CIRs (Dumbovic et al. 2012; Maričić et al. 2014). FD is one of the beneficial effects of solar activity in that the impact of GCRs on Earth is moderated by Earth-directed CMEs. Belov et al. (2014) investigated FDs making use of the CME database from SOHO and GCR intensity from the worldwide neutron monitor network. They found good correlations of the FD magnitude with the CME initial speed, the ICME transit speed, and the maximum solar wind speed. Full halo CMEs showed the maximum FD, followed by partial halos and non-halos. Figure 14 shows that faster and wider CMEs are more effective in causing FDs. Note that the CMEs in the 360º bin are most effective in causing FD. Full halo CMEs generally originate close to the disk center and head directly toward Earth and hence are effective in producing FDs and geomagnetic storms. These results are consistent with the findings by Abunina et al. (2013), who found the solar sources of disturbances causing the maximum FD are close to the central meridian (E15 to W15).

Despite large international efforts in understanding FDs, there is still a lot to learn. The current model of FDs consisting of two-step decrease has recently been questioned. It is not clear if only a subset of CMEs originating from the disk center is effective in causing FDs (Jordanova et al, 2012). However, the study of FDs has been gaining interest in recent times because of the space weather applications. For example, the development of Global Muon Detector Network (GMDN – Munakata et al. 2005) has greatly enhanced the possibility of forecasting ICME arrival using the network (see e.g., Rockenbach et al. 2014 for a review).

### 3.2.9 Spatial Structure of CMEs

Even before the discovery of white-light CMEs, the concept of magnetic loops from the Sun driving shocks was considered (Gold 1962). In Gold's picture, a magnetic bottle from the Sun drives a fast magnetosonic shock which stands at certain distance from the bottle. Such a shock was first identified by the Mariner 2 mission in 1962 (Sonett et al. 1964). Koomen et al. (1974) identified white-light CMEs with the Gold bottle. Burlaga et al. (1981) confirmed the basic picture of Gold using in-situ data by identifying the shock, sheath and the driving magnetic structure. Near the Sun, MHD shocks were inferred from metric type II radio bursts for several decades ago (see e.g., Nelson and Melrose, 1985). The overall CME structure consisting of a flux rope enclosing a prominence core and driving a shock outside has been considered by theorists a while ago (e.g., Kuin and Martens 1986),but it took another two decades before the white-light shock structure of CMEs was observed in coronagraphic images (Sheeley et al. 2000). A recent study based on coronagraph observations concluded that a flux rope structure can be discerned in ~40% of CMEs observed near the Sun (Vourlidas et al. 2013).

**3.2.10 White-Light and EUV Signatures of CME-driven Shocks**

Fast forward interplanetary shocks (hereafter simply called "shocks") are driven by either fast CMEs or high speed streams. So far, no "blast wave" shocks have been detected in the interplanetary medium by spacecraft instrumentation. Shocks compress and heat the upstream plasma and magnetic fields (Kennel et al., 1985). Thus, the immediate downstream (or sheath) region may be visible at times. Shocks form from a steepening of magnetosonic waves. To identify whether a wave is a shock or not, it must be shown to have a supermagnetosonic speed in its normal direction. Methods of analyses can be found in Tsurutani and Lin (1985) and the geoeffectiveness of shocks and discontinuities in Tsurutani et al. (2011b).

There have been several recent studies on white-light shocks (Vourlidas et al. 2003; Michalek et al. 2007; Gopalswamy et al. 2008a; Gopalswamy 2009; Gopalswamy et al. 2009b; Ontiveros and Vourlidas 2009; Bemporad and Mancuso, 2011; Gopalswamy and Yashiro 2011; Maloney and Gallagher 2011; Kim et al. 2012; Poomvises et al. 2012) that have provided a better understanding of the CME structure beyond the classical three-part structure (Hundhausen et al. 1987). The availability of STEREO and SDO observations increased our ability to visualize the CME-shock system and understand the shock formation and coronal plasma properties.

The dome structure surrounding newly erupted CMEs has been recognized as the three-dimensional counterpart of the so-called EIT waves (Patsourakos and Vourlidas, 2009; Veronig et al. 2010; Ma et al. 2011; Kozarev et al. 2011; Warmuth and Mann 2011; Gallagher and Long 2011; Harra et al. 2011; Gopalswamy et al. 2012b; Selwa et al. 2013; Temmer et al. 2013; Liu and Ofman 2014; Nitta et al. 2013a). The wave nature of EUV waves was also established based on the fact that they are reflected from nearby coronal holes (Long et al. 2008; 2013; Gopalswamy et al. 2009c; Olmedo et al. 2012; Shen et al. 2013; Kienreich et al. 2013: Kwon et al. 2013). Gopalswamy and Yashiro (2011) estimated the coronal magnetic field within the SOHO coronagraphic field of view (6–23 Rs) using the fact that the standoff distance of the white-light shock with respect to the radius of curvature of the driving flux rope is related to the shock Mach number and the adiabatic index (Russell and Mulligan 2002; Savani et al. 2012). Since the shock speed is measured from the coronagraphic images, these authors were able to derive the Alfven speed and magnetic field in the ambient medium. Poomvises et al. (2012) extended this technique to the interplanetary medium and showed that the derived magnetic field strength is consistent with the HELIOS in-situ observations. This technique will be extremely important to compare future in-situ observations from missions to the Sun such as Solar Orbiter and Solar Probe Plus, currently under development (Müller et al. 2013). The standoff-distance technique was also applied to a CME-shock structure observed by SDO on 2010 June 13, which showed that the technique can work as close to the Sun as 1.20 Rs, where the shock first formed (Gopalswamy et al. 2012b; Downs et al. 2012). The shock formation heights derived from SDO/AIA and STEREO/EUVI have provided direct confirmation that CME-driven shocks have enough time to accelerate particles to GeV energies from a height of ~1.5 Rs before they are released when the CME reaches a height of about 3-4 Rs (Gopalswamy et al. 2013a,b; Thakur et al. 2014). The low shock formation heights applies only to those CMEs, which quickly accelerate and attain high speeds (see e.g., Bein et al. 2011).

### 3.2.12 Shocks Inferred from Radio Observations

Type II radio bursts in the metric domain, traditionally observed from ground based observatories indicate shock formation very close to the Sun (e.g., Kozarev et al. 2011; Ma et al. 2011; Gopalswamy et al. 2012b). Imaging these bursts provide important information such as the magnetic field in the ambient medium (Hariharan et al. 2014). These bursts indicate the height of shock formation in the corona as evidenced by EUV shocks and Moreton waves (see e.g. Asai et al. 2012a,b). Radio emission from interplanetary shocks in the form of type II bursts provide important information of shock propagation in the heliosphere (Gopalswamy 2011). CMEs with continued acceleration beyond the coronagraph field of view (FOV) may form shocks at large distances where they become super-magnetosonic (faster than the upstream magnetosonic wave speed). Shocks forming at large distances of the Sun may or may not produce type II radio bursts (Gopalswamy et al 2010c). Radio-quiet CMEs (those lacking type II radio bursts) typically have positive acceleration in the coronagraphic field of view and become super-magnetosonic in the interplanetary (IP) medium at large heliocentric distances. Deceleration of radio-loud CMEs near the Sun and the continued acceleration of radio-quiet CMEs into the IP medium make them appear similar at 1 AU. However, there is a better chance that radio-loud CMEs produce an energetic storm particle event (Mäkelä et al. 2011) and strong sudden commencement/sudden impulse (Veenadhari et al. 2012), suggesting that stronger shocks near the Sun do matter. In fact Vainio et al. (2014) have shown that the cut-off momentum of particles observed at 1 AU can be used to infer properties of the foreshock and the resulting energetic storm particle event, when the shock is still near the Sun.

By combining STEREO/HI (Heliospheric Imager) observations, interplanetary radio bursts observations, and in-situ measurements from multiple vantage points, Liu et al. (2013) showed that it is possible to track CMEs and shocks. In particular, they were able to study CME interaction signatures in the radio dynamic spectrum. The drift rate of the type II radio bursts can also be converted into shock speed for comparison with the CME speed derived from HI observations, providing a method to predict shock arrival (e.g. Xie et al. 2013a).

### 3.2.13 Shock Critical Mach Numbers

There is renewed interest in shock critical Mach numbers and their evolution with heliocentric distance (Gopalswamy et al. 2012b; Bemporad and Mancuso 2011; 2013; Vink and Yamazaki 2014). For example, some radio-loud shocks may dissipate before reaching 1 AU (Gopalswamy et al. 2012c) indicating that the Mach number is dropping to 1 or below. Bemporad and Mancuso (2011) concluded that the supercritical region occupies a larger surface of the shock early on, but shrinks to the nose part of the shock as it travels away from the Sun. Vink and Yamazaki (2014) introduced a different critical Mach number ($M_{acc}$), which is substantially larger than the first critical Mach number ($M_{crit}$) of quasi-parallel shocks (Kennel et al., 1985), but similar to $M_{crit}$ of quasi-perpendicular shocks. According to these authors, the condition $M_{acc} > \sqrt{5}$ seems to be required for particle acceleration, which may be relaxed when seed particles exist.

The clear identification of an interplanetary magnetic cloud (MC) with a CME by Burlaga et al. (1982) replaced the Gold bottle by a flux rope. Although the MC definition by Burlaga et al. (1982) was narrower than the flux rope definition (magnetic field twisted around an axis), the terms "flux rope" and "MC" are interchangeably used after Goldstein (1983) showed that the MC magnetic field can be modeled by a flux rope. All theories of CME eruption and propagation use the flux rope as the fundamental structure in their calculations, either preexisting or formed during eruption (Yeh, 1995; Chen 1997; Riley et al. 2006; Forbes et al. 2006; Chen 2012; Kleimann 2012; Lionello et al. 2013; Janvier et al. 2013; Démoulin 2014). Extensive CME observations from the SOHO mission have helped perform many studies on CME flux ropes. Chen et al. (1997) showed that the observed CME structure in the LASCO field of view can be interpreted as the two-dimensional projection of a three-dimensional magnetic flux rope with its legs connected to the Sun.

The CME flux rope is thought to be either pre-existing or formed out of reconnection during the eruption process and is observed as an MC in the interplanetary medium (see e.g., Gosling, 1990; Leamon et al., 2004; Qiu et al., 2007). On the other hand, it is possible that a set of loops from an active region on the Sun can simply expand into the interplanetary (IP) medium and can be detected as an enhancement in the magnetic field with respect to the ambient medium (Gosling, 1990) without any flux-rope structure. The in-situ magnetic signatures will be different in the two cases. A spacecraft passing through the flux rope center will see a large, smooth rotation of the magnetic field throughout the body of the interplanetary CME (ICME), while the expanded loop system will show no rotation. If we take just the IP observations, we may be able to explain MCs as flux ropes and non-MCs as expanding loops. However, they should show different charge-state characteristics (see e.g. Aguilar-Rodriguez et al. 2006; Gopalswamy et al. 2013c) because of the different solar origins. The flux rope forms during the flare process and hence is accessed by the hot plasma resulting in high charge states inside MCs when observed at 1 AU. Expanding loops, on the other hand, should not have high charge states because no reconnection is involved. Riley and Richardson (2013) analyzed Ulysses spacecraft measurements of ICMEs and concluded the ICME may not appear as MCs because of observing limitations or the initiation mechanism at the Sun may not produce MCs.

In a series of two coordinated data analysis workshops (CDAWs) a set of structure of CMEs, 54 CME-ICME pairs were analyzed to study the flux-rope nature of CMEs (see Gopalswamy et al. 2013d for the list of papers based on these CDAWs). It was found that MCs and non-MCs were indistinguishable based on their near-Sun manifestations such as white-light CMEs and flare post-eruption arcades. In particular, the CMEs were fast and the flare arcades were well defined (Yashiro et al. 2013). Fe and O charge states at 1 AU were also indistinguishable between MCs and non-MCs, suggesting a similar eruption mechanism for both types at the Sun (Gopalswamy et al. 2013c). Combined with the fact that CMEs can be deflected towards or away from the Sun-Earth line (Gopalswamy et al., 2009d), the observing geometry (i.e., the observing spacecraft may not cross the flux rope axis) seems to be the primary reason for the non-MC appearance of flux ropes (see e.g., Kim et al. 2013). Many authors have advocated that all ICMEs are flux ropes (Marubashi 2000; Owens et al. 2005; Gopalswamy 2006a), but the single point observations at 1 AU may miss it. Marubashi et al. (2015) showed that almost all ICMEs can be fit to a flux rope if a locally toroidal flux rope model is considered in addition to the cylindrical flux rope model. Similarly, the active region helicity and the helicity of the ICMEs were in good agreement (Cho et al. 2013).

Using SDO/AIA data, Zhang et al. (2012) reported that flux ropes exist as a hot channel before and during an eruption. The structure initially appeared as a twisted and writhed sigmoid with a temperature as high as 10 MK, and then transformed into a semi-circular structure during the slow-rise phase, which was followed by a fast acceleration and flare onset. Cheng et al (2013) reported that the hot channel rises before the first appearance of the CME leading front and the flare onset of the associated flare. These results indicate that the hot channel acts as a continuous driver of the CME formation and eruption in the early acceleration phase. Li and Zhang (2013a,b) reported on the eruption of two flux ropes from the same active region within 25 minutes of each other on 2012 January 23. The two flux ropes initially rose rapidly, slowed down, and accelerated again to become CMEs in the coronagraph FOV. The two CMEs were found to be interacting in the coronagraph FOV as observed by SOHO and STEREO (Joshi et al. 2013). Li and Zhang (2013c) also studied homologous flux ropes from active region 11745 during 2013 May 20–22. All flux ropes involved in the eruption had a similar morphology.

**3.2.14 Propagation Effects: Deflection, Interaction and Rotation of CMEs**
Once a CME is ejected from the Sun, its 3-D geometry at a far-away location such as Earth depends on the evolution in the changing background solar wind and magnetic field (Temmer et al. 2011). The CME flux ropes expand, so the magnetic content typically decreases. CMEs can be deflected in the latitudinal and longitudinal directions by pressure gradients (magnetic + plasma). The CME flux ropes can also be distorted by changing flow speeds in the background. Finally, CMEs may also rotate, so the orientation of the magnetic field inferred from solar observations may not match what is observed at Earth. It is possible that many of these effects can occur simultaneously (Nieves-Chinchilla et al. 2012). At present there are a few techniques to connect CMEs observed at the Sun with their interplanetary counterparts. Interplanetary type II bursts detected by Wind/WAVES and STEREO/WAVES instruments can track CME-driven shocks all the way from the Sun to the observing spacecraft located at 1 AU (Xie et al. 2013a). Interplanetary scintillation (IPS) observations track turbulence regions surrounding CMEs (typically the sheath region) also over the Sun-Earth distance (see e.g., Manoharan 2010; Jackson et al. 2013). The heliospheric imagers on board STEREO track CMEs in white light over the Sun-Earth distance (e.g. Möstl and Davies 2013).

A combination of CME tracking in the inner heliosphere using STEREO Heliospheric Imagers (HIs) and numerical simulations have greatly enhanced CME propagation studies. The interaction of CMEs with the ambient medium is the primary propagation effect. This interaction is represented by an aerodynamic drag that dominates beyond the coronagraphic field of view. Close to the Sun, the propelling force and gravity dominate (see e.g. Vrsnak and Gopalswamy 2002). Defining the background is one of the key inputs needed for understanding CME propagation (see e.g., Roussev et al. 2012; Arge et al. 2013). However, there are other processes that can significantly affect the propagation of CMEs: CME-CME interaction (see e.g. Gopalswamy et al. 2001a; 2012c; Temmer et al. 2012; Harrison et al. 2012; Lugaz et al. 2012; Liu et al. 2014a; Sterling et al. 2014; Temmer et al. 2014) and CME deflection by large-scale structures such as coronal holes and streamers (Gopalswamy et al. 2009d, 2010; Shen et al. 2011; Gui et al. 2011; Wood et al. 2012; Kay et al. 2013; Panasenco et al. 2013; Gopalswamy and Mäkelä 2014).

Different types of interaction become predominant during different phases of the solar cycle. During the rise phase, when polar coronal holes are strong and CMEs originate at relatively higher latitudes, the polar coronal holes are effective in deflecting CMEs (e.g., Gopalswamy et al. 2008b). During the maximum phase, CMEs occur in great numbers, so CME-CME interaction is highly likely (e.g., Gopalswamy et al. 2012c; Lugaz et al. 2013; Chatterjee and Fan 2013; Farrugia et al. 2013; Kahler and Vourlidas, 2014). CME interactions can also result in CME deflection and merger (Shen et al. 2012). In the declining phase, low-latitude coronal holes appear frequently, so CME deflection by such coronal holes becomes important (Gopalswamy et al. 2009d; Mohammed et al. 2012; Mäkelä et al. 2013). The deflections are thought to be caused by the magnetic pressure gradient between the eruption region and the coronal hole (Gopalswamy et al 2010d; Shen et al 2011; Gui et al 2011).

Determining the initial orientation of flux ropes has been possible by fitting a flux rope to coronagraph observations (Thernisien 2011; Xie et al. 2013b). Such fitting already provides a lot of information on the deviation of CME propagation direction from the radial (e.g., Gopalswamy et al. 2014b). Isavnin et al (2014) defined such initial flux rope orientation using extreme-ultraviolet observations of post-eruption arcades and/or eruptive prominences and coronagraph observations. Then they propagated the flux rope to 1 AU in a MHD-simulated background solar wind and used in-situ observations to check the results at 1 AU. They confirmed that the flux-rope deflection occurs predominantly within 30 Rs, but a significant amount of deflection and rotation happens between 30 Rs and 1 AU. They also found that that slow flux ropes tend to align with the streams of slow solar wind in the inner heliosphere.

**3.2.15 CME Arrival at Earth**

There have been many attempts recently to convert the knowledge gained on CME propagation to predict the arrival times at 1 AU. The CME travel time essentially depends on the accurate estimate of the space speed of CMEs and the background solar wind speed, these have been estimated based on single view (SOHO) as well as from multiple views (SOHO and STEREO). Davis et al. (2010) found that CME speeds derived from STEREO/COR2 and Thernisien (2011) forward-fitting model were in good agreement, although CME speeds changed in the HI FOV depending on the near-Sun speed. Millward et al. (2013) developed the CME Analysis Tool (CAT), which models CMEs to have a lemniscate shape, which is similar to ice-cream cone model. They showed that the leading-edge height and half-angular width of CMEs can be determined more accurately using multi-view data. Colaninno et al. (2013) tracked 9 CMEs continuously from the Sun to near Earth in SOHO and STEREO images and found that the time of arrival was within ±13h. Gopalswamy et al. (2013e) considered a set of 20 Earth-directed halos viewed by SOHO and STEREO in quadrature, so as to obtain the true earthward speed of CMEs. When the speeds were input to the Empirical Shock Arrival (ESA) model, they found that the ESA model predicts the CME travel time within about 7.3 h, which is similar to the predictions by the ENLIL model. They also found that CME-CME and CME-coronal hole interaction can lead to large deviations from model predictions. Vršnak et al. (2014) compared the arrival-time predictions from the "WSA-ENLIL+Cone model" and the analytical "drag-based model" (Vršnak et al. 2013) and found that the difference in predictions had an absolute average of 7.1 h. Compared with observations, the drag-based model had an average absolute difference of 14.8 h, similar to that for the ENLIL model (14.1 h). Xie et al. (2013a) compared travel times of CMEs when ENIL+ cone model and ENLIL+ flux rope model were used. They found that the ENLIL+ flux rope model results showed a slight improvement (4.9 h vs. 5.5h). They also found that predictions based on kilometric type II bursts improved significantly when the ENLIL model density was used rather than the average solar wind plasma density in deriving shock speeds from the type II drift rate. The improvement was typically better by ~2h. Möstl et al. (2014) derived the absolute difference between predicted and observed ICME arrival times for 22 CMEs as 8.1 h (rms value of 10.9 h). Empirical corrections to the predictions reduced arrival times to within 6.1 h (rms value of 7.9 h). Echer et al. (2010) attempted to identify the solar origins of the November 2004 superstorms on Earth using existing interplanetary propagation routines published in the literature. They found that during highly active solar intervals, the predictions were sometimes ambiguous, in agreement with the comments above. Thus there has been a steady progress in predicting the arrival time of CMEs, which need to be continued and expended to the prediction of Bz values, which are crucial to predict the strength of geomagnetic storms.

**3.3. CMEs and Geomagnetic Storms**

One of the direct consequences of CMEs arriving at Earth's magnetosphere is the geomagnetic storm. The primary link between a geomagnetic storm and a CME is the out of the ecliptic component (Bz) of the interplanetary magnetic field (Gonzalez et al., 1994; Zhang et al., 2007; Echer 2008a,b, 2013). Echer et al. (2008a) conclusively showed that for all 90 major (Dst < -100 nT) storms that occurred during cycle 23, it was the Bz component that was responsible for the storms (some people have thought that it was possible that the IMF By component was also important). When Bz is negative (south pointing), then the CME field reconnects with Earth's magnetic field (Dungey, 1961) causing the geomagnetic storm. While the Bz component is negligible in the quiet solar wind, CMEs contain Bz by virtue of their flux rope structure. Fast CMEs drive shocks, so the compressed sheath field between the flux rope and the shock can also contain Bz (Tsurutani et al., 1988). Thus both the flux rope and sheath can be source of Bz and hence cause geomagnetic storms. One of the common indicators of the strength of geomagnetic storms is the Dst index (expressed in nT), which is computed as the horizontal component of Earth's magnetic field measured at several equatorial stations (now a SYM-H index is available which is essentially a one min resolution Dst index). Yakovchouk et al. (2012) reported significant difference between the local and global peak storm intensities: the local storm minima were found to be 25-30% stronger than the global minima. Here we consider only global peak intensities. Major storms have Dst $\leq$ -100 nT and are mostly caused by CMEs. Gonzalez et al. (2007) and Echer et al. (2008a) studied all 90 storms with Dst $\leq$-100 nT for cycle 23. They divided ICMEs proper from their upstream sheaths. They found that roughly half of the storms that were caused by CME/sheaths were due to CMEs and half due to sheath fields. Yakovchouk et al. (2012) found that 10% of major storms are caused by CIRs. Echer et al. (2008a) determined that 13% of the cycle 23 storms were caused by CIRs.

It should be noted that all studies of "superstorms" or storms with Dst intensities < -250 nT have been caused by magnetic clouds (Tsurutani et al., 1992; Echer et al., 2008b). Such intense storms are not caused by sheaths or CIRs. One might ask why not? The argument is a simple one. The slow solar wind magnetic field is ~5-7 nT. It has been shown by Kennel et al. (1985) that fast shocks can compress the magnetic field by a maximum factor of ~4, regardless of the shock Mach number. Thus interplanetary sheath fields and CIR magnetic fields should have maximum field strengths of ~20 to 35 nT. In contrast, magnetic cloud fields have been 50-60 nT and in exceptional cases ~100 nT. Thus even if the sheath and CIR magnetic fields are totally southward they are small by comparison to MC fields.

There is one exception to this above explanation. In cases of ARs where there are multiple CMEs and multiple shocks, the shocks can "pump up" sheath magnetic field intensities. This has been shown to be the case for both CAWSES intervals of study (Tsurutani et al., 2008, 2013).

One of the early signatures of the weak solar cycle 24 has been the drastically reduced number of major (Dst ≤ -100 nT) geomagnetic storms (Echer et al. 2011b; 2012; Gopalswamy, 2012; Richardson 2013; Kilpua et al. 2014) ¬. A plot of the Dst index as a function of time in Fig. 15 shows that the frequency and amplitude of the storms in cycle 24 are the lowest in the space age (cycles 19–24). There were storms with Dst < -200 nT in every cycle since 1957, except in cycle 24, in which the storms never exceeded a strength of 140 nT. Several historical storms (including the recent ones on 1989 March 14 with Dst = -589 nT and on 2003 November 20 with Dst = -422 nT) can be found in the plot (see Cliver et al 1990; Gopalswamy et al. 2005b,c).

Figure 16 compares the number of major storms as a function of time for cycles 23 and 24, including the last couple of events of cycle 22. We see that the gap between the last storm of cycle 22 and the first storm of cycle 23 was only ~16 months. On the other hand, the gap was about 4 times larger between cycles 23 and 24 as the Sun emerged out of the deep solar minimum following cycle 23. The number of storms in the corresponding phases of cycles 23 and 24 were 37 and 11, respectively, indicating a 70% reduction of major storms in cycle 24. This is considerably more than the reduction of the number of fast and wide CMEs from the disk center or the overall reduction in the number of energetic CMEs (<30%, see Table 1). Gopalswamy et al (2014a) showed that the reduction in the heliospheric total pressure (plasma + magnetic) makes CMEs expand more in cycle 24 thereby reducing their density and magnetic content that ultimately result in weaker storms. This can be understood from the empirical relation between the Dst and the product VBz of the interplanetary structure causing the storm:

Dst = -0.01V|Bz| – 32 nT (Wu and Lepping 2002; Gopalswamy 2010).

Table 2 shows the list of major storms from cycle 24 (updated to the end of 2014 June 30 from Gopalswamy, 2012). There were only 12 major storms with the Dst ranging from -103 nT to -137 nT. Two storms with the Dst close to -100 nT are also included in Table 2. All storms were due to CMEs, except one caused by a CIR (2013 June 1). The average Dst of the 11 CME storms was only -119 nT. Over the corresponding epoch in cycle 23, there were 36 storms with an average Dst of -158 nT. Thus the number of major storms in cycle 23 was at least three times more and 33% stronger. The average speed of the CMEs in Table 2 is 1025 km/s, compared to 722 km/s for the corresponding epoch in cycle 23. i.e., cycle-24 CMEs producing major storms were faster than those in cycle 23 by 42%. The halo fraction was comparable between the two cycles (7 out of 11 or 64% in cycle 24 compared to 19 out of 31 or 61% in cycle 23). In other words, cycle-24 CMEs need be faster to produce storms similar to the ones in cycle 23.

Six of the eleven CME storms in Table 2 were due to southward Bz in the ICME sheaths. These include one event in which the sheath was actually a preceding magnetic cloud: the shock from a disk-center CME on 2014 Feb 16 from S11E01 associated with a M1.1 flare at 09:20 UT entered into a preceding fully-south (FS) cloud. The source of the magnetic cloud itself is not clear, but most likely a faint CME associated with an eruption in the northwest quadrant on 2014 Feb 14 around 02:33 UT (a faint CME at 4:28 UT CME was barely discernible in LASCO images). The shock compressed the preceding CME and enhanced the Bz that caused the storm. This is a good example that ICMEs can be affected by shocks from other eruptions. In the remaining five storms, southward Bz was in the cloud portion. Even cycles such as cycle 24 are supposed to have more north-south (NS) clouds (the ones with leading northward Bz). However, there was only one such cloud in Table 2. All others were of FS clouds, which are high-inclination clouds with south-pointing axial field. Note that the Bz values ranged from -11.1 nT to -28.7 nT with an average of -18.5 nT. The Bz values in cycle 24 were in a narrower range compared to those in cycle 23.

**3.4 Flares and the Ionosphere**

It has long been known that solar flares create sudden ionospheric disturbances or SIDs (Thome and Wagner, 1971; Mitra, 1974). However the extreme intensity of the Halloween flares and the rise of using global positioning systems (GPS) for ionospheric research has allowed major advances to be made in flare ionospheric research (Tsurutani et al. 2005; Afraimovich et al., 2009). Now with ground based receivers virtually everywhere on Earth (with ocean coverage still a bit of a problem), high time resolution, global coverage is now possible.

Figure 17 shows the change in the ionospheric total electron content (TEC) during the peak of the October 28, 2003 solar flare, from 1100 to 1108 UT. The flare peak time was taken from the unsaturated SOHO SEM narrow band EUV detector. The NOAA GOES X-ray detector was saturated. Thomson et al. (2004) has estimated its strength of the flare as large as X45 ± 5 via other means. A quiet day background of 27 October was subtracted from the 28 October data to get this difference plot in Figure 17. The subsolar point (Africa) is at the center of the figure. The data points are individual ground observations of GPS satellites.
The solar flare causes the largest TEC enhancement at the subsolar region with a TEC enhancement of 22 TEC units. The nightside region shows no TEC change, as expected. This is the largest ionospheric TEC change due to a solar flare ever detected.

Figure 18 shows the simultaneous onset of the October 28 flare (both at SOHO and GOES), the Libreville ionospheric TEC enhancement and the dayglow enhancement at ~1100 UT. The SEM data (unsaturated) shows a double peak structure. The ionospheric TEC rose from ~1100 UT to ~1105 UT and then less rapidly from 1105 to ~1118 UT, where a peak value of ~25 TEC above background was attained. One hour prior to the flare, the background TEC was 82 TECU so the flare caused a ~30% increase in the ionospheric content in this region. This is the largest flare-TEC event on record. The dayglow (from the TIMED GUVI O 135.6 nm and N2 LBH 141.0 to 152.8 nm band) increased most rapidly from ~1100 to ~1104 and peaked at ~1115 UT. The TEC enhancement lasted far longer (~3 hrs) than the flare itself (~20 min). The cause is that the EUV portion of the flare causes photoionization at altitudes above ~170 km where the recombination time scale is hours (Tsurutani et al., 2005).

**3.5 CMEs and the Ionosphere**
The southward magnetic component of ICMEs (and their upstream sheaths) create magnetic storms, which are enhancements in the Earth's outer radiation belts. The magnitude is measured by ground based magnetometers near the equator giving the Dst and SYM-H indices. CMEs/magnetic storms also cause severe ionospheric effects as well. Energetic particle precipitation into the auroral zones lead to local heating and neutral atmospheric expansion called the "disturbance dynamo" (Blanc and Richmond, 1980; Scherliess and Fejer, 1997). It has also been noted that the interplanetary electric field reaches ionospheric levels (Nishida 1968; Kelley et al., 2003) causing other effects. More recently, it was noted that CME interplanetary electric fields penetrated down to the equatorial ionosphere and lasted for hours (Tsurutani et al., 2004). The electric fields were "unshielded", contrary to theoretical expectations. The unshielded storm-time electric fields lead to what is called the "daytime superfountain effect" illustrated in Fig. 19.

Figure 19 show the "dayside superfountain effect" for the October 30, 2003 Halloween magnetic storm. The CHAMP satellite pass before the storm (blue trace), shows the two equatorial ionization anomalies (EIAs) located at ~ ±10°. With time, the ionosphere and EIAs are uplifted to higher magnetic latitudes and have higher intensities. In the first pass after storm onset (red curve), the EIAs have peak intensities of ~200 TECU at ~± 20° MLAT. In the following pass, a peak intensity of ~330 TECU is detected at ~30° MLAT. The cause of this remarkable feature is the interplanetary dawn-dusk electric field which uplifts the upper ionosphere by E x B convection (the Earth's magnetic field is aligned in a north-south direction at the magnetic equator). As the electrons and ions are convected to higher altitudes and latitudes, solar irradiation replaces the uplifted plasma by photoionization, leading to an overall increase in the TEC (Tsurutani et al., 2004; Mannucci et al., 2005).

### 3.6. Coronal holes and CIRs

Coronal holes play a number of important roles in Sun-Earth connection. Polar coronal holes indicate the strength of the polar field and hence the level of solar activity in the following cycle (Gopalswamy et al. 2012a; Selhorst et al. 2011; Shibasaki 2013; Mordvinov and Yazev 2014; Altrock 2014). Coronal holes in the equatorial region are good indicators of imminent high-speed streams (HSS) and CIRs arriving at Earth (Tsurutani et al., 1995, 2006; Cranmer 2009; Verbanac et al. 2011; Akiyama et al. 2013; Borovsky and Denton 2013). The empirical relationships established between HSS characteristics and the related geomagnetic activity provides an advance warning of impending CIR storms (Tsurutani et al., 2006; Verbanac et al. 2011). Coronal holes also deflect CME-driven shocks and CMEs that have important space weather consequences (Gopalswamy et al. 2009d; 2010; Olmeda et al. 2012; Kay et al. 2013; Mäkelä et al. 2013). The deflection by coronal holes can be so large that CMEs originating from close the disk center of the Sun do not arrive at Earth while the shocks do. Coronal-hole deflection may also make the shock and the driving flux rope appear unaligned (Wood et al.

2012). Both high-speed streams and CIRs result in various types of magnetospheric responses (Tsurutani et al. 2006; Denton and Borovsky 2012; Borovsky and Denton 2013). Coronal holes also seem to play a critical role in deciding whether CMEs originating at latitudes >30º can produce ground level enhancement events (Gopalswamy and Mäkelä 2014), although their effect on large SEP events has not been conclusive (Kahler et al. 2014). In this section we provide an overview of the recent progress on the geomagnetic response.

**3.6.1 CIRs, CIR Storms and HSS geomagnetic activity**

High-speed streams (HSSs) originate from coronal holes and form CIRs when they collide with the slower solar wind ahead (see e.g. Smith and Wolf, 1976; Gosling 1996). The compressed interaction region has a higher density and temperature and the magnetic field intensities and fluctuations are amplified. When the field within the CIR contains southward Bz, geomagnetic activity ensues (Borovsky 2013). However the fields within CIRs are typically highly fluctuating (compressed Alfven waves). Thus the character is considerably different from that of the magnetic fields within magnetic clouds. CIRs typically do not cause magnetic storms with Dst < -100 nT (Tsurutani et al., 1995, 2006; Echer et al. 2008a). The geomagnetic activity is typically in the range -50 nT < Dst < -100 nT. Geomagnetic storms caused by CMEs and CIRs (plus the following HSSs) differ in some important ways (Denton et al. 2006; Jardonova et al. 2012; Liemohn and Katus 2012; Verbanac et al. 2013; Borovsky and Denton 2013; Keese et al. 2014). For example, the CMEs and CIRs (plus HSSs) have different magnetospheric response resulting in different development of various current systems and geomagnetic activity within the Earth's magnetosphere and ionosphere (Mannucci et al., 2005, 2008, 2012; Thayer et al, 2008; Lei et al., 2008, 2011; Verkhoglyadova et al., 2011, 2013, 2014; Verbanac et al. 2013). Keesee et al. (2014) performed superposed epoch analysis of 21 CME-driven and 15 CIR-driven storms during the June 2008-April 2012 time frame and different evolution of the ion temperature: the ion temperature increased in the recovery phase of CIR storms, while it increased rapidly at the onset of CME storms and cooled off during the main phase. Borovsky and Denton (2013) compared CIR storms associated with helmet streamers and pseudo-streamers. They found that pseudo-streamer CIR storms tend not to have a calm (Tsurutani et al., 1995) before the storm, with weaker superdense plasma sheet and electron radiation belt dropout.

**3.6.2 Effect of the Weak Solar Activity**

Denton and Borovsky (2012) compared the magnetospheric effects of 93 strong (maximum speed ~600 km/s) and 22 weak (maximum speed ~500 km/s) high-speed streams. The weak HSSs were observed during the extended minimum that followed cycle 23. The strong HSSs were from earlier periods. A superposed epoch analysis showed that the solar wind velocity, in combination with the southward component of the IMF, largely governs the magnetospheric response to HSSs. In particular, the ring current was stronger and the magnetospheric electron flux was higher in the strong HSSs (see Fig. 20). They point out that the difference in the evolution of particle flux can be attributed to the physical conditions in the magnetosphere that differ significantly under the two types of HSSs.

Even though the CIRs were weak during the prolonged minimum (see e.g. Echer et al. 2011a), they had interesting effects on the ionosphere and atmosphere. The ionospheric response to the weak CIRs was marginal but observable (Araujo-Pradere et al. 2011). However, the weak recurrent geomagnetic activity due to CIRs did produce distinct variability in the thermospheric density at an altitude of 400 km above ground (Lei et al. 2011 and references therein). The thermosphere was found to respond globally with the density varying by ~75%. Most importantly, they were able to isolate the effect of geomagnetic activity from the EUV forcing because the EUV flux remained roughly constant during these CIR intervals. Thermospheric density variations also showed the periodicities in CIRs due to the spatial distribution of low-latitude coronal holes on the Sun, as did the ionosphere.

Hajra et al. (2013) studied a subset of high speed streams, those that had particularly high geomagnetic activity associated with them, called High-Intensity Long-Duration Continuous Geomagnetic Activity (HILDCAA: Tsurutani and Gonzalez, 1987). These events are defined by occurring outside magnetic storms (thus in the HSS proper), lasting at least 2 days and having a peak AE > 1,000 nT. The AE could not drop below 200 nT for longer than 2 hrs. Hajra et al. (2013) found that the HILDCAAs ordered high speed streams quite well. The solar cycle dependence of these events for 3 ½ solar cycles is shown in Fig. 21.

Figure 21 gives the solar cycle dependence of HILDCAAs. HILDCAAs are most often detected during the declining phase of the solar cycle, but they can be detected during of the other phases as well.

Figure 22 gives the E > 0.6, > 2.0 and > 4.0 MeV electron fluxes detected at geosynchronous orbit. All of the 38 HILDCAA events that occurred during 1995 to 2008 (cycle 23) are included in this superposed epoch analysis. The start time is the onset of the HILDCAAs. For all of the events, the relativistic electron fluxes at the peak of the HILDCAA/HSS events were higher than their pre-event values. Thus it is clear that acceleration had taken place and not simple removal and replacement of the electrons. It was also found that 100% of the events had simultaneous electromagnetic chorus waves present (when data were available) for the $5 < L < 10$ and $00 < MLT < 12$ region where chorus is expected to occur (Tsurutani and Smith, 1977; Meredith et al., 2001, 2003). The figure shows that the E > 0.6, > 2.0 and > 4.0 MeV electrons occurred ~1.0 day, ~1.5 day and ~2.5 days after HILDCAA onsets. The results are in good agreement with the theoretical predictions (Horne and Thorne 2003), Horne 2007) The ~10-100 keV substorm injected electrons generate chorus by the loss cone/temperature anisotropy instability. These same waves cyclotron resonate with the ~100 keV electrons to produce relativistic electrons on ~day time scales. In this scenario, the E = 0.6 MeV electrons are accelerated first, then the E = 2.0 MeV population from the 0.6 MeV electrons and so forth. This "boot strap" acceleration scenario was the interpretation of Hajra et al. (2014) for the various delay times of the relativistic electrons shown in the figure.

The overall scenario of the magnetospheric relativistic electron acceleration starts at the Sun (Tsurutani et al. 2006; 2010). Supergranular circulation is the source of the Alfven waves (Hollweg, 2006). These Alfven waves are carried from the coronal holes at the Sun to the Earth by the HSSs. The southward components of the Alfven waves lead to magnetic reconnection and the geomagnetic activity indicated in the HILDCAAs. The injection of ~10 to 100 keV electrons by substorms/convection events within the HILDCAAs lead to chorus wave growth and the chorus accelerate electrons to relativistic energies. All features of this scenario have now been confirmed, except for the Alfven wave source. Perhaps the Solar Probe Plus or Solar Orbiter will be able to finally confirm this last part of the scenario.

### 3.6.3 CIR and HSS Ionospheric and Atmospheric Effects

Verkhoglyadova et al. (2011) studied ionospheric and atmospheric effects of HSSs in the solar minimum interval, 25 March to 26 April 2008. This was the study interval for the Whole Heliospheric Interval (WHI) science team. As stated previously this was near the solar minimum but about a year before the geomagnetic minimum.

Figure 23 top panel shows the TIMED SABER enhanced NO infrared radiation coming from the atmosphere during the WHI HILDCAA activity (third panel) during the magnetic storm recovery phases (bottom panel). These are the intervals of high AE activity. Seven wide latitude bins are indicated. Most of this irradiation is coming from high latitudes, presumably the auroral zone region.

Mlynczak et al (2003) views NO irradiation as a "natural thermostat". Energy input into the upper atmosphere during high AE intervals is converted to heat and changes the distribution of NO in the thermosphere and its radiative properties. The radiation from NO cools the atmosphere. NO is believed to account for ~50% of the estimated energy input to the atmosphere from the magnetic storm.

The ionospheric TEC effects during CIRs/HSSs are weak but are clearly present (Verkhoglyadova et al., 2011). There is enhanced TEC both at low latitudes (second panel: -30° < MLAT < +30°) and at middle latitudes (third panel: 40° to 60° MLAT). The middle latitude disturbance is most likely due to particle precipitation in the auroral zone. However the equatorial disturbances are less well understood. It may be associated with a disturbance dynamo effect or the dayside equatorial superfountain. More study is warranted.

### 3.6.4 Microwave Enhancement in Coronal Holes and Solar Wind Speed

Akiyama et al. (2013) also considered weak and strong CIRs but within solar cycle 23 (1996–2005). The weak and strong CIRs resulted in the Dst index > -100 nT and < -100 nT, respectively. The CIRs were associated with low-latitude coronal holes identified in EUV images from SOHO/EIT. They measured the area of the coronal hole in EUV and in microwaves (from the Nobeyama Radioheliograph). Coronal holes appear bright in microwaves and the area of microwave enhancement overlaps with that in EUV but not completely. The area of microwave enhancement is typically a third of the area observed in EUV CH on the average (Gopalswamy et al. 2000). There are at least two reasons for this difference: (1) the microwave enhancement originates in the chromosphere, so the area is expected to be smaller due to magnetic field expansion, and (2) microwave enhancement is an indicator of higher magnetic field within the coronal hole, which is usually patchy within the coronal hole. A clear correlation between the CH size and the solar wind speed is well known (e.g. Nolte et al. 1976).

Figure 24 shows a linear correlation between the maximum solar wind speed and the CH area in EUV ($r = 0.62$) and microwaves ($r = 0.79$). The correlation is slightly better for the CH area in microwaves, suggesting that faster wind originates from higher magnetic field regions. Figure 24 also shows that the intense storms (Dst < -100 nT) are associated with faster winds compared to the weaker storms (Dst > -100 nT). This may simply be due to the fact that the Dst index depends on the product of solar wind speed and the southward component of the CIR magnetic field.

### 3.6.5 The Lone Major CIR storm of Cycle 24

The only major CIR storm of cycle 24 (as of this writing) occurred on 2013 June 1. The source of the storm at the Sun was a low-latitude coronal hole (Fig. 25). The storm was relatively intense (Dst = -119 nT), with the strength similar to that of CME storms of cycle 24. The coronal hole was relatively large and straddled the equator as can be seen from the SDO/AIA image taken in the beginning of 2013 May 30. An outline of the coronal hole superposed on the SDO/HMI magnetogram shows that it encloses a region of positive magnetic polarity. In-situ observations by spacecraft at L1 show that the CIR had a large magnetic field (~25 nT) and the Bz was also large (~-21 nT), which was responsible for the geomagnetic storm.

### 3.7. Large SEP events and GLE events

Solar energetic particles (SEPs) are part of the mass emission from the Sun, intimately connected to CMEs and flares. The current paradigm is that the material in the corona and IP medium is accelerated by CME-driven shocks to produce the large SEP events (e.g., Reames 1999; 2013). Detailed investigation on the connection between CME-driven shocks and SEPs became possible only after the advent of SOHO coronagraphs that routinely imaged CMEs in the coronal region from where the energetic particles are released. In particular, the highest energy (GeV) particles have been studied in relation to CMEs only for the past couple of solar cycles (Cliver 2006, Gopalswamy et al. 2012d; Nitta et al. 2012; Mewaldt et al. 2012; Miroshnichenko et al. 2013). Studies on the connection of SEPs to CMEs have been further enhanced by the extended coverage provided by STEREO in a number of ways, including observations all around the Sun. The Heliospheric Imagers and the inner coronagraph COR1 have extended the spatial domain over which CMEs are observed. In addition, EUV observations from STEREO/EUVI and SDO/AIA have improved our ability to study the early phase of CMEs (see e.g., Aschwanden et al. 2014). For example, the coronal height where shocks form can be readily determined from EUV and COR1 images, so that the time available for accelerating SEPs can be estimated accurately. Characterizing the size of large eruptions from behind the limb has been made possible by estimating the soft X-ray fluxes based on the known correlation between EUV flux and soft X-ray flux for frontside events (Nitta et al. 2013b).

Even though STEREO was launched by the end of 2006, no large SEP events occurred until the second half of 2010 in the CAWSES-II period. From then on, there were many SEP events that have been studied extensively. The STEREO particle detectors observed SEP events from off the Sun-Earth line and, together with detectors at Sun-Earth L1, provided information on the longitudinal distribution of SEP intensity (e.g., Dresing et al., 2012; Rouillard et al. 2012; Mewaldt et al., 2013). It also became possible to study the radial dependence of SEP intensities, thanks to the observations provided by MESSENGER spacecraft (Lario et al. 2013). Observations from multiple spacecraft have also been used to test diffusive shock acceleration: Wang and Yan (2012) performed a dynamical Monte Carlo simulation of the CME-driven shock that occurred on 2006 December 14 using an anisotropic scattering law and found that the simulated results of the shock's fine structure, particle injection, and energy spectrum were in good agreement with the observations.

### 3.7.1 Longitudinal Dependence of SEP Intensity

Lario et al. (2006; 2013) were able to fit a Gaussian to the ensemble of multispacecraft SEP peak intensities as a function of the longitudinal distance between the solar source (inferred from flare observations) and the magnetic connection point for each spacecraft on the Sun. They found that such distributions are slightly offset to the west from the solar source. Long-lived particle injection from CME-driven shocks has been suggested as the reason for this shift (Lario et al. 2014). They also estimated that the shock height is within ~40 Rs when the particle injection from the shock maximizes.  Given the observation that CME-driven shocks form at a heliocentric distance of 1.5 to 5 Rs (Tsurutani et al., 2003; Gopalswamy et al. 2013b), these findings indicate that observations close to the Sun are most important for large SEP events. Furthermore, the range of heliocentric distances matches with the frequency range of IP type II bursts (Gopalswamy et al. 2012c).

### 3.7.2 SEP Intensity Variability

Attempts to understand the variability in SEP intensities have been focused on two aspects: (1) Source factors involving CME properties, (2) Environmental factors such as ambient magnetosonic speed, preceding CMEs, deflection by large-scale structures, seed particles, and turbulence. The CME speed and width are the primary source parameters. In cycle 23 about ~75% of SEP-producing CMEs were found to be halos (Gopalswamy et al. 2006b) for the whole cycle as well as for the first 62 months. On the other hand, 100% of SEP-producing CMEs are halos in cycle 24 so far. This remarkable result can be understood in terms of the anomalous expansion of CMEs in cycle 24 due to the reduced heliospheric pressure. The average sky-plane speed of SEP-producing CMEs is ~1500 km/s, slightly higher than that in cycle 23 over the corresponding phase of the solar cycle (Gopalswamy 2012). These results lead to the conclusion that the cycle-24 CMEs need to have more kinetic energy to produce SEP events similar to those in cycle 23. In other words, the cycle-24 CMEs seem to be less efficient in accelerating particles.

### 3.7.3 SEP Events Associated with Weak and Strong Eruptions

SEP events are generally associated with large soft X-ray flares, but this may not indicate a physical connection (Cliver et al 2012). In fact, soft X-ray flare size is not a good indicator of SEP association. Gopalswamy et al. (2014b) investigated 59 major eruptions with flare size $\geq$M5. The CME flux rope location was on the front side of the Sun for 55 of them. Only 20 (or 36%) of the 55 eruptions were associated with large SEP events including those detected by STEREO-B (located behind the east limb). On the other hand, out of the 31 large SEP events detected by GOES during the first 62 months of cycle 24, ten were associated with <M5.0 flares (half of them associated with C-class flares). SEP-associated CMEs in both these populations (size $\geq$M5.0 and <M5.0) were very fast: the average speeds were 2300 and 1720 km/s, respectively. This result is consistent with the idea that energetic particles in large SEP events are primarily accelerated by CME-driven shocks, as was pointed out by Cliver (2006) and Gopalswamy et al. (2012d) for GLE events. This point is further illustrated in Figure 26, which shows the source locations of the 55 major eruptions ($\geq$M5.0) of cycle 23 and the ones with 20 large SEP events. The eruptions are divided into two groups according to their speed ($\geq$1500 km/s for 28 events and <1500 km/s for 27 events). Out of the 28 fast CMEs, 17 (or 61%) were associated with large SEP events, compared to only 3 (or 11%) of the 27 slower CMEs. Figure 26 also shows the solar sources of SEP events associated with the 10 weaker eruptions (M<5.0): only 2 (or 20%) of the weak eruptions were associated with <1500 km/s CMEs while 8 (or 80%) were associated with $\geq$1500 km/s CMEs. The faster CMEs without SEP events mostly occurred when there was high background level of SEPs due to previous events. In one case, the CME was jet-like (the width was <<60°).

**3.7.4 Solar Cycle Effects**

Unlike major geomagnetic storms, the number of large SEP events did not decrease significantly in cycle 24. Figure 27 shows the number of large SEP events binned into Carrington rotations from September 1995 (end of cycle 22) to April 2014 (middle of cycle 24). The gap between the last event of cycle 22 and the first event of cycle 23 was ~2 years. The corresponding gap was nearly doubled during the cycle 23/24 minimum. However, once the activity started, the number of large SEP events occurred roughly at the same rate as in the first five years of cycle 23. As of June 2014, there were 34 large SEP events in cycle 24 compared to 47 over the corresponding phase in cycle 23. The reduction is only by 26%, similar to the reduction in the number of fast and wide western CMEs (W20–W90) in cycle 24 (Gopalswamy et al. 2014a).

The peak intensity and fluence of the >10 MeV SEP events in cycle 24 were not too different from the corresponding values in cycle 23 (Fig. 28). This is in stark contrast to the major reduction in the intensity and number of major geomagnetic storms in cycle 24. The SEP behavior can also be explained in terms of the altered state of the heliosphere: as Fig. 13 shows, the Alfven speed of the corona decreased during the rise phase of cycle 24 compared to the corresponding phase in cycle 23. Lower upstream magnetosonic speeds result in higher Mach number shocks for a given CME speed, increasing the likelihood of the occurrence of large SEP events. The two solar cycles, however, differ drastically in the case of higher-energy SEP events. The number of large SEP events with >500 MeV particles declined by 58%, while the number of GLE events declined by 71%. These cannot be explained by the 22% drop in the number of fast and wide CMEs in cycle 24. This issue is addressed in the next section.

It should be noted that the largest SEP events are associated with quasi-parallel shocks (Kennel et al. 1984a,b). This is because of the presence of upstream turbulence ahead of the shocks (Tsurutani et al., 1983), leading to enhanced Fermi-type acceleration across the shocks.   .

**3.7.5 The Paucity of GLE Events in Cycle 24**

Ground level enhancement (GLE) in SEPs represent the highest energy particles accelerated by the Sun, making the particles penetrate Earth's atmosphere to the troposphere where they produce air showers like those produce by the galactic cosmic rays (GCRs) and the air showers are detected by ionization chambers, muon detectors, and neutron monitors on the ground. These hard-spectrum events affect navigation systems, spacecraft electronics and operations, space power systems, manned space missions, and commercial aircraft operations (Shea and Smart 2012; Mewaldt et al. 2012; Kudela 2013; Ruffolo et al. 2013). In particular they can be a significant radiation exposure to humans in space and in air-planes on polar routes. GLE events also have important implications for VLF wave propagation: Zigman et al. (2014) found that at energies up to ~2 GeV the ionization rate for solar protons may exceed the GCR ionization by 1.5 orders of magnitude.

Typically about a dozen GLEs occur in each solar cycle, which is about 15% of the number of large SEP events during cycles 19–23 (Shea and Smart 2008). However, there were only two GLEs during cycle 24 so far, even though there were 34 large SEP events, amounting to <6% (Gopalswamy et al. 2013a; Thakur et al. 2014; Papaioannou et al. 2014). The paucity of GLE events in cycle 24 cannot be explained by the 22% reduction in the number of fast and wide CMEs originating in the traditional GLE longitudes (W20–W90). A combination of several factors is needed to explain the drastic reduction in the number of GLE events: (1) The reduction in the ambient magnetic field in the corona reduces the efficiency of shock acceleration especially for GeV particles, even though it is not a problem in accelerating ~10 MeV particles. (2) Many CMEs have nonradial motion either due to the coronal environment or inherent asymmetry in the source region. The nonradial motion makes the shock nose not well connected to Earth (Gopalswamy et al. 2013a; Gopalswamy and Mäkelä 2014). This means that even if GeV particles were accelerated at the shock nose, they may not reach the observer. (3) The CME may be ejected into a locally tenuous corona making the shock weaker.

Figure 29 compares the 2014 January 6 GLE CME (Thakur et al. 2014) with a non-GLE CME on January 7 that was even faster than the GLE CME. The nose of the January 6 CME was in the ecliptic; it was well below the ecliptic for the January 7 CME. In other words, the nose of the January 6 shock was magnetically connected to Earth, while just the flank of the shock was connected during the January 7 event. The 2014 January 6 CME resulted in a GLE even though the speed was smaller than the January 7 CME (1700 km/s vs. 3100 km/s). The January 6 CME originated from behind the limb (S02W102) while the January 7 CME originated from close to the disk center (S15W29). The nonradial motion in CMEs can be attributed to deflection by nearby coronal holes (Gopalswamy et al. 2009d; 2010; Kay et al. 2013) or to the inherent asymmetry in the distribution of magnetic field strength in the source region (see e.g. Sterling et al. 2011). In the source region of the January 7 CME, the flare ribbons were located to the south of intense fields that did not participate in the eruption, but might have deflected the CME to the south. There were also a coronal hole to the northeast of the eruption region, which might have also contributed to the deflection.

The fact that both CMEs in Fig. 29 were large SEP events raises an important question on the location of particle acceleration on the shock surface. The stringent requirement of the ecliptic distance for GLE CMEs suggests that GLEs may be accelerated at the shock nose, where the shock is the strongest. On the other hand, magnetic connection to any part of the shock seem to suffice for large SEP events. This means that lower energy SEPs are accelerated over most of the shock surface. This is also consistent with the study by Dalla and Agueda (2010) who found that the probability of detecting SEP events remained constant up to a latitude of 28º. Gopalswamy and Mäkelä (2014) examined the latitudinal connectivity issue in historical GLE source regions that were at latitudes >30º. They were able to show that in all the higher latitude cases, there was a polar coronal hole, a streamer, or a pseudo-streamer poleward of the GLE source region suggesting deflection toward equator and hence enhancing the possibility of shock nose connection to Earth.

The requirement that the CME nose be in the ecliptic for GLE events may also have implications for the shock geometry. It is reasonably certain that GLEs are released when the CME is at a height of ~3 Rs. Since the shock nose has the largest heliocentric distance compared to other parts, it is likely that the nose is above the source surface and the upstream field magnetic field is open. This suggests that GLE-producing regions of the shock may have quasi-parallel geometry. Recent investigations confirm the suggestion that CME-driven shocks are likely to be supercritical and quasi-parallel near the nose, while subcritical and quasi-perpendicular at the flanks (Bemporad et al 2014; Bemporad and Mancuso 2011; 2013). It is well known that the first critical Mach number is the smallest (~1.5) for quasi-parallel shocks and the largest for quasi-perpendicular shocks (~2.7) (see e.g. Mann et al. 2003). Supercritical shocks are known to accelerate particles more efficiently (Burgess 2013). Thus for a given shock speed (~2000 km/s for GLE events) and the typical Alfven speed of ~600 km/s at 3 Rs (Gopalswamy et al. 2001b), the quasi-parallel shocks are likely to be definitely supercritical, while the quasi-perpendicular shocks may or may not be supercritical. These conclusions on the shock geometry and connectivity need to be further explored and modeled for a full understanding of particle acceleration by CME-driven shocks.

**3.8. Extreme Space Weather**

Given that CMEs are responsible for the largest geomagnetic storms and SEP events, it is natural to think that some properties of CMEs or their source regions may result in extreme events. If we define extreme events as those, which lie on the tail of a distribution, we can readily identify the CME's speed as one critical parameter. From the cumulative distribution of CME speeds, one infers that there are not many CMEs with speeds exceeding ~3500 km/s (e.g. Gopalswamy et al 2010b). When a CME has a speed >3500 km/s, it may be thought of as an extreme event. The consequences of such an energetic CME are likely to be extreme also. An extremely fast CME will drive a shock, which will accelerate particles to very high energies. If all conditions for the acceleration of particles are conducive (low ambient magnetosonic speeds, good connection of the shock nose to the observer, a quasi-parallel shock, and a preconditioned ambient medium) one might expect an extreme SEP event. Kovaltsov and Usoskin (2014) determined the cumulative occurrence probability distribution of SEP events based on directly-measured SEP fluences for the past 60 years, estimates based on the terrestrial cosmogenic radionuclides $^{10}$Be and $^{14}$C for Holocene timescale, and cosmogenic radionuclides measured in lunar rocks on a timescale of up to 1 Myr. They concluded that SEP events with a > 30 MeV proton fluence greater than $10^{11}$ (protons cm$^{-2}$ yr$^{-1}$) are not expected on a Myr timescale.

The SEP event depends just on the outer structure of a fast CME, viz., the shock. It should be noted that the most intense interplanetary shock detected at 1 AU distance from the sun was detected on March 8 in the 7-17 March 2012 CAWSES II study interval and had a magnetosonic Mach number of ~9.4 (Tsurutani et al., 2014). However the maximum shock Mach number possible, assuming an ICME speed of 2700 km/s is 45 (Tsurutani and Lakhina, 2014). One speculation to explain the difference is that it is possible that the acceleration of energetic particles at the shock leads to "shock damping", reducing the shock intensity. Theoretical studies are needed to confirm/deny this hypothesis.

On the other hand, the generation of a large geomagnetic storm depends on the internal structure of the CME, especially the magnetic cloud (MC, Burlaga et al., 1982) and filament (Kozyra et al. 2014) internal to it. What is important is that the magnetic field intensities are high and southward. It has been shown empirically that fast CMEs have particularly intense magnetic fields (Gonzalez et al., 1998). Sheath magnetic fields (Tsurutani et al., 1988) are swept up slow solar wind fields and not part of the CME proper. Sheath fields can have intensities of only ~4 times the upstream ambient field strength (Kennel et al., 1985; Tsurutani and Lakhina, 2014) unless multiple shocks are compressing the sheath, so this region is, in general, unimportant for the cause of extreme storms. If the magnetic cloud has its axis pointing to the south and the magnetic field is high, one could expect strong magnetic interconnection between the interplanetary magnetic field and the Earth's magnetopause magnetic field (Dungey, 1961) and therefore an intense geomagnetic storm. In addition if the shock remained strong, it would produce an extreme impulse on the magnetosphere that can expose geosynchronous orbit to the solar wind (Tsurutani and Lakhina, 2014) and a sudden (positive) impulse ($SI^+$) on the ground (Tsurutani et al., 2008). An intense shock would also produce an extreme energetic storm particle (ESP) event at Earth (Tsurutani et al. 2009). Thus for solar terrestrial relationships, an extreme event in its origin will also have extreme consequences at Earth and near-Earth space environment.

Tsurutani and Lakhina (2014) considered an extreme scenario for geomagnetic response. Since the CME speed and its magnetic content can ultimately be traced to the solar source (Gopalswamy 2010), an extreme CME would be born in an active region with enormous free energy. Going by the highest field strength ever observed in a sunspot (~6100 G, Livingston et al. 2006) and the largest active region area (5000 msh, Newton 1955), one can estimate a free energy of ~$10^{36}$ erg. This is two to three orders of magnitude larger than the amount of free energy estimated for NOAA AR 10486, which produced several of the Halloween CMEs (Gopalswamy et al. 2005c). If the entire ~$10^{36}$ erg goes into the kinetic energy of a CME of mass ~$10^{18}$ g, then the CME speed would be ~14000 km/s, about 4 times larger than the CME speed ever observed.

Recently, Maehara et al. (2012) investigated 365 stellar flares with energies in the range $10^{33}$ to $10^{35}$ erg using the Kepler mission data. Fourteen of those flares were from Sun-like stars (G-type main sequence stars with rotational periods >10 days and surface temperature in the range 5600–6000K). They estimated that superflares with energy $10^{35}$ erg occur once in 5000 years. Note that this energy is an order or magnitude less than the maximum free energy estimated by Gopalswamy et al. (2010). It must be pointed out that CMEs are not observed directly in stellar eruptions, but if the eruption mechanism is similar to that on the Sun, one would expect even higher CME kinetic energies than the $10^{35}$ erg flare energy found by Maehara et al. (2012). In fact, the CME kinetic energy gains the largest share among various ways in which the free energy in an active region is divided (Emslie et al. 2012). However one should take note that the amount of energy that goes into the flare particles is deducted from the CME energy (Tsurutani and Lakhina, 2014).

Shibata et al. (2013) also investigated theoretically the possibility of superflares on the Sun using the current dynamo model, and concluded that it is indeed possible to generate magnetic flux necessary for producing superflares of energy in the range $10^{34}$ - $10^{35}$ erg within the next 40 years. The Solar Evolution and Extrema (SEE) project of the VarSITI (Variability of the Sun and Its Terrestrial Impact) program of SCOSTEP is expected to make further progress on this issue over the next few years.

The Carrington Flare of 1859 remains one of the important benchmark events for extreme solar events. The flare itself was detected with naked eyes, indicating that it is certainly an unexpected event. Cliver and Dietrich (2013) estimated that the soft X-ray flare size of X45, compared to X35 for the cycle-23 flare on 2003 November 4. The Carrington eruption did produce a historical geomagnetic storm, the highest intensity on record: Dst = -1760 nT (Tsurutani et al. 2003). Kataoka (2013) estimated that the probability of a Carrington-type geomagnetic storm occurring within the next decade is ~4–6%. Yermolaev et al. (2013) performed a statistical analysis of the OMNI data for the period 1976–2000 and concluded that a Carrington-type event could occur once every 500 years (see also Alves et al. 2011). From the flare onset to the geomagnetic storm onset, Carrington (1859) gave a time of ~17.5 hours from the flare to the storm, which indicates a CME speed of ~2360 km/s (Gopalswamy et al. 2005c).   The all-time record for the shortest CME transit time from the Sun to 1 AU was the August 12, 1972 event (Vaisberg and Zastenker, 1976) which was ~14.6 hours. With such a fast transit time and large magnetic cloud magnetic field, why didn't a large magnetic storm occur? It was discovered that the magnetic field orientation within the magnetic cloud was northward, so the Earth had geomagnetic quiet rather than a large storm (Tsurutani et al. 1992).   It has been shown that for northward IMFs impinging on the Earth's magnetosphere, one has extreme geomagnetic quiet rather than an intense storm (Tsurutani et al. 1992; Du et al., 2011). These CME transit times are relatively close to the maximum time of 12 hrs, assuming an initial CME speed of 3,000 km/s (Tsurutani and Lakhina, 2014). Two of the Halloween storms in cycle 23 had such short transit times: 18.9 and 19.7 hours (Tsurutani et al., 2005; Mannucci et al., 2005) for the 2003 October 28 and 29 CMEs, respectively. In the extremely weak cycle 24, there was another extreme event on 2012 July 23 observed by multiple spacecraft. This well-observed event merits some additional discussion.

**3.8.1 July 23 2012 event**

On July 23, 2012 the GOES satellite reported a large SEP event with a peak >10 MeV flux of ~ 12 pfu (1 pfu = 1 particle sr$^{-1}$ cm$^{-2}$ s$^{-1}$), a very ordinary event. However, the solar source of the associated CME was ~45° behind the west limb and the CME was heading roughly towards STEREO-A (located at W121), which observed the CME as a full halo. STEREO-A also detected a 5000-pfu SEP event with an ESP event ~10 times larger, similar to the 1992 March 23 and 1989 October 20 events (43,000 and 40,000 pfu, respectively). The CME arrived at STEREO-A in ~19 h, making it one of the historical events potentially greater than the Halloween 2003 events (Russell et al. 2013b; Baker et al. 2013; Mewaldt et al. 2013; Gopalswamy et al. 2014b). The STEREO-A SEP event started ~4 h earlier than the GOES event and ~8 h earlier than the STEREO-B event because STEREO-A was better connected to the source than the other two spacecraft. There are a number of similarities between the 2012 July 23 event and the 20 October 1989 event studied by Lario and Decker (2002).

If the CME were heading toward Earth instead of STEREO-A, it would have caused another geomagnetic storm of historical proportions (Liu et al. 2014b). The total magnetic field strength was ~109 nT in the shock sheath and ~60 nT in the ICME. The southward component had a peak value of -52 nT. Liu et al. (2014b) estimated that the Dst index to be in the range -1150 nT to -600 nT. The simple empirical relation reported by Gopalswamy (2010) also gives a high Dst index: Dst = -0.01VBz – 32 nT. Plugging in V=1500 km/s and Bz = 52 nT, the Dst index becomes -812 nT. This is half of the strength reported for the Carrington storm (-1760 nT, Tsurutani et al. 2003) and a quarter of the maximum possible of Dst = -3500 nT (Tsurutani and Lakhina). That is if the magnetosphere does not saturate at Dst = -2500 nT as Vasyliunas (2011) has argued. If we compare with the estimate of Siscoe et al. (2006), who used hourly averages to estimate the Dst index as -850 nT, we see that the July 2012 storm has the same strength as the Carrington event (see also Cliver and Dietrich 2013 who estimate Dst ~-900 nT). The important point is that such extreme storms can occur even during the subdued heliospheric conditions prevalent in cycle 24. It is possible that the July 2012 storm already represent a reduction in Bz because of the anomalous CME expansion; otherwise the storm could have been even bigger than the current estimates. If the ICME were south pointing with the 109 nT field during the July 2012 event (Liu et al. 2014b) and the speed were similar to the shock speed (2250 km/s), the empirical formula (Dst = 0.01VBz – 32 nT) would predict a storm of ~ -2500 nT.

### 3.8.2 1859 Carrington Ionospheric Event

Scientists now understand the dayside superfountain effect well and can model such events. Figure 30 shows a simulation of the Carrington magnetic storm effects in the dayside ionosphere. The Huba et al (2002) SAMI-2 code was used which describes the dynamics and chemical evolution of seven ion species and seven corresponding neutral species. The code solves collisional two-fluid equations for electrons and ions along dipole magnetic fields taking into account photoionization of neutrals, recombination of ions and electrons and chemical reactions. The code was modified to allow an external electric field input (Verkhoglyadova et al. 2008) and was recently further modified to insert 3 hr Ap indices instead of daily values. A 20 mV/m electric field for a duration of 1 hr was assumed (Tsurutani et al., 2003; 2011).

In panel a), prior to the event, the two EIAs are clearly noted one spanning ~ -5° to -30° and a second from ~ +5° to 20° LAT (we use a 3.25 x $10^6$ ions/$cm^{-3}$ to define the enhancement areas). Panel c shows the oxygen peaks are located at ~530 to 920 km and ~500 to 900 km for the southern and northern regions, respectively. The magnetic latitude ranges are ~-20° to -35° LAT and +30° to +45° LAT. The peak values reach ~6 x $10^6$ $O^+$ $cm^{-3}$ at altitudes centered at ~700 km. It is noted that the densities of the uplifted EIA peaks at 850 km and 1000 km were ~4 x $10^6$ and ~3.5 x $10^6$ $cm^{-3}$, respectively. These latter ion densities are substantially greater than quiet time neutral density values, being ~40 times at 850 km and ~300 times at 1,000 km. Thus low altitude satellite drag will be substantially increased in such an extreme storm.

## 4. Conclusions

The weak activity as the Sun climbed towards the maximum phase of solar cycle 24 has provided an enormous opportunity to study the Sun and the heliosphere under quieter conditions. The space weather is less severe compared to cycle 23 in terms of intense geomagnetic storms and powerful SEP events. On the other hand, the cosmic-ray intensity remains exceptionally high after reaching the highest levels in the space age during the extended solar minimum. The geomagnetic ap index was the lowest on record. The latter was not only due to the changes of the Sun but also the location of the Earth relative to solar active regions (coronal holes). The vast array of ground and space based instruments has helped define this altered state of the Sun and heliosphere. The CAWSES-II program enabled many researchers to take the systems approach and be mindful of the implications of their research for solar terrestrial relationship. From the eruptive events during the weak solar cycle to the dynamo problem and the grand minima of solar activity, the Sun has triggered a plethora of research activities, including the MiniMax24 activity that recognizes the importance of studying weak activity. The VarSITI program will address the declining phase of the weak cycle 24, when there will be phase-specific phenomena such as frequent CIRs and high-speed streams from the Sun.

The TG3 activities of the CAWSES-II program will naturally connect to the International Studies of Earth-affecting Transients (ISEST) project of VarSITI. Campaigns are being conducted in observing transient solar events under the MiniMax24 component of the ISEST project. The working groups under ISEST will deal with the theory, modeling, and detailed analysis of the campaign events. There will also be a close connection to the Solar Evolution and Extreme (SEE) project in examining flares on sun-like stars and extreme events such as the Carrington event and the 2012 July 23 event. These connections provide continuity to the SCOSTEP science activities and make progress without serious interruptions.

**List of abbreviations used**

AIA, Atmospheric Imaging Assembly

AU, Astromomical Unit

CAWSES-II, Climate and Weather of the Sun-Earth System II Phase

CAT, CME Analysis Tool

CDAW, coordinated data analysis workshop

CH. coronal hole

CIR, Corotating interaction region

CME, Coronal mass ejection

EUV, Extreme-ultraviolet

FD, Forbush decrease

FOV, field of view

GCR, galactic cosmic rays

GLE, Ground level enhanncemet

GOES, Geostionary Operational Environment Satellite

HI, Heliospheric Imager

HILDCAA, High-Intensity Long-Duration Continuous Geomagnetic Activity

HSS< high speed stream

ICME, Interplanetary coronal mass ejection

IP, Interplanetary

LASCO, Large Angle and Spectrometric Coronagraph

MC, magnetic cloud

MHD, Magnetohydrodynamics

SCOSTEP, Scientific Committee On Solar Terrestrial Physics

SDO, Solar Dynamics Observatory

SEPs, Solar energetic particles

SGD, Solar Geophysical Data

SOHO Solar and Heliospheric Observatory

SSN, Sunspot number

STEREO, Solar Terrestrial Relations Observatory

TEC, total electron content

TG3 Task Group 3 (of CAWSES II)

TSI, Total solar irradiance

## Competing interests

The authors' declared that they have no competing interest. All institutional and national guidelines for the care and use of laboratory animals were followed.

## Authors' contributions

NG created the overall structure of the review article and was approved by the co-authors. He was responsible for drafting the full manuscript.    BT contributed to several sections, especially on the geospace aspects. YY wrote some sections in the manuscript and coordinated the effort to check all the references.    All authors read and approved the final manuscript.

## Authors' information

NG is a NASA scientist at the Goddard Space Flight Center and the current President of SCOSTEP. As the President, he was responsible for the successful completion of the CAWSES-II program to which the present review belongs. NG is interested in the study of the origin, propagation, and Earth impact of solar disturbances. BT is a JPL scientist heavily involved in space plasma physics and the geospace impact of solar disturbances. YY is a radio astronomer from the National Astronomical Observatory of China interested in solar physics and space weather.

**Acknowledgements**

The authors thank Kazunari Shibata and Joseph Borovsky for providing leadership to the Task Group 3: How does short term solar variability affect the geospace environment? The authors also thank CAWSES-II leaders Susan Avery, Alan Rodger, Toshitaka Tsuda, and Joseph Davila for their effort in running a successful SCOSTEP scientific program. We thank P. Mäkelä and Jing Huang for carefully reading the manuscript. We also thank Kazunari Shibata for his comments on the manuscript. The work of NG was supported by NASA's LWS TR&T program. Portions of this research was performed at the Jet Propulsion Laboratory, California Institute of Technology under contract with NASA. We thank the anonymous referees for the helpful comments.

Table 1. Comparison between CME numbers in solar cycles 23 and 24

| CME Property | Cycle23[a] | Cycle 24 | Ratio |
|---|---|---|---|
| All CMEs | 5086 (89.2/mo) | 8201 (134.44/mo) | 1.51 |
| W<30° | 1732 (30.38/mo) | 3907 (64.05) | 2.11 |
| W≥30° | 3354 (58.84/mo) | 4294 (70.39/mo) | 1.20 |
| W≥60° | 1858 (32.6/mo) | 2205 (36.15/mo) | 1.13 |
| W = 360° | 178 (2.99/mo) | 199 (3.06/mo) | 1.02 |
| V≥900 km/s & W≥60° | 189 (3.32/mo) | 142 (2.33/mo) | 0.70 |
| ≥C3.0 flares, limb | 273 (4.7/mo) | 214 (3.45/mo) | 0.73 |

[a]Over the same epoch as cycle 24; [b]Ratio of cycle-24 rate to cycle-23 rate

**TABLE 2** Major geomagnetic storms of cycle 24 (Dst < -100 nT)

| # | Date and Time of Storm | Dst (nT) | CME Onset | V[a] km/s | W[b] deg | Eruption Location | Bz (nT) Loc.[d] |
|---|---|---|---|---|---|---|---|
| 1 | 20110806 04:00 | -107 | 08/04 04:12 | 1315 | H | N19W36 | -21.4 Sh |
| 2 | 20110927 00:00 | -103 | 09/24 12:48 | 1915 | H | N15E58 | -28.7 Sh |
| 3 | 20111025 02:00 | -137 | 10/22 01:25 | 593 | H | N40W30[c] | -19.9 Sh |
| 4 | 20120309 09:00 | -133 | 03/07 01:36 | 1825 | H | N17E27 | -17.7 FS |
| 5 | 20120424 05:00 | -107 | 04/19 15:12 | 540 | 142 | S30E71[c] | -15.3 FS |
| 6 | 20120715 18:00 | -125 | 07/12 16:48 | 885 | H | S15W01 | -18.6 FS |
| 7 | 20121001 04:00 | -133 | 09/28 00:12 | 947 | H | N06W34 | -20.0 Sh |
| 8 | 20121009 09:00 | -111 | 10/05 02:48 | 612 | 284 | S23W31[c] | -16.2 FS |
| 9 | 20121114 08:00 | -108 | 11/09 15:12 | 559 | 276 | N08W19[c] | -18.9 NS |
| 10 | 20130317 21:00 | -132 | 03/15 07:12 | 1063 | H | N11E12 | -11.1 Sh |
| 11 | 20130601 09:00 | -119 | CIR | ---- | ---- | CH | -21.4 IR |
| 12 | 20140219 09:00 | -112 | 02/14 04:28 | ???? | ??? | S12W25 | -15.5 FS[e] |

[a]CME speed with an average value of 1021 km/s   [b]CME width (H = Full halos: 7/11 or 63%). [c]Disappearing Solar Filament (DSF) events with the centroid of the filament given.  [d]Origin of negative Bz: Sh = sheath; FS = fully south (axis of the high inclination cloud pointing south); NS = north south cloud (negative Bz in the rear end of the cloud); IR=interface region between fast and slow winds.   [e]The sheath contains a preceding FS MC.

**Figure legends**

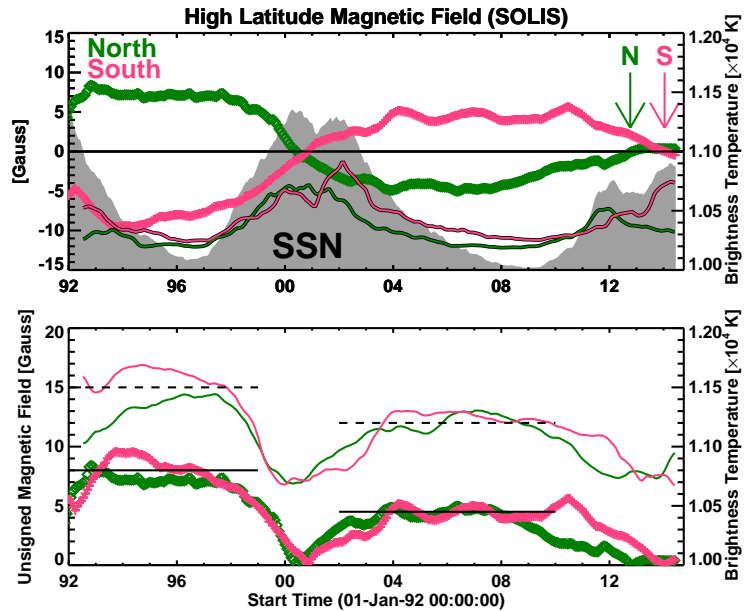

**Figure 1.** Comparison between solar cycles 23 and 24 from various observations: (top) An overview of solar cycle 24 with respect to cycle 23 using three sets of observations: the international sunspot number (SSN in gray), polar magnetic field strength in the north and south polar regions (B, averaged over the latitude range 60–90º, thick lines), and the low-latitude microwave brightness temperature (Tb, averaged over 0–40º, thin lines). The low-latitude Tb is due to active regions. N and S point to the time when the polar magnetic fields vanished before the sign reversal. (bottom) Polar microwave brightness temperature (Tb, averaged over the latitude range 60–90º) and the unsigned polar magnetic field strength (B). The horizontal dashed and solid lines roughly indicate the average levels of Tb and B, respectively for the two cycles. The time of vanishing polar B occurs first in the north and then in the south during cycle 23 and 24. However, the lag is more pronounced in cycle 24. Note that the maximum phase has ended in the northern hemisphere (indicated by the steady increase in polar Tb by the end of 2013). All quantities are smoothed over 13 Carrington rotations.

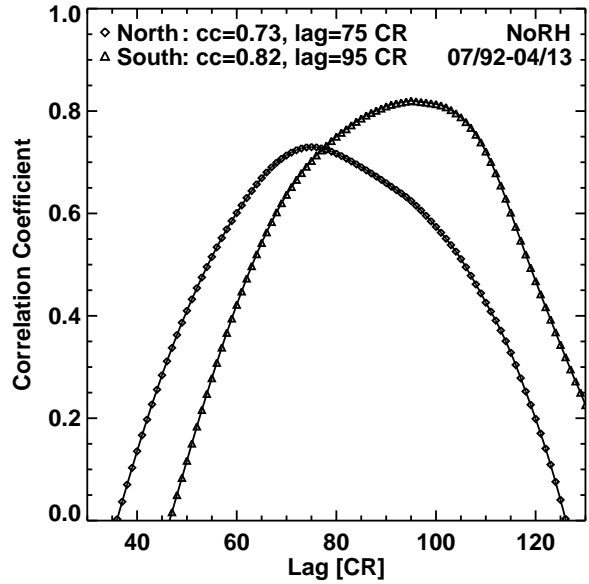

**Figure 2.** Polar field of cycle 23 and activity strength in cycle 24: Correlation between the polar microwave brightness (proxy to the poloidal field strength of the Sun) of one solar cycle and the low-latitude brightness (proxy to the solar activity) of the next as a function of the time lag in units of Carrington rotation (CR). The correlation is quite high and supports the flux transport dynamo model. The correlation coefficients are shown on the plot for the northern and southern hemispheres along with the lag extent (in number of CRs).

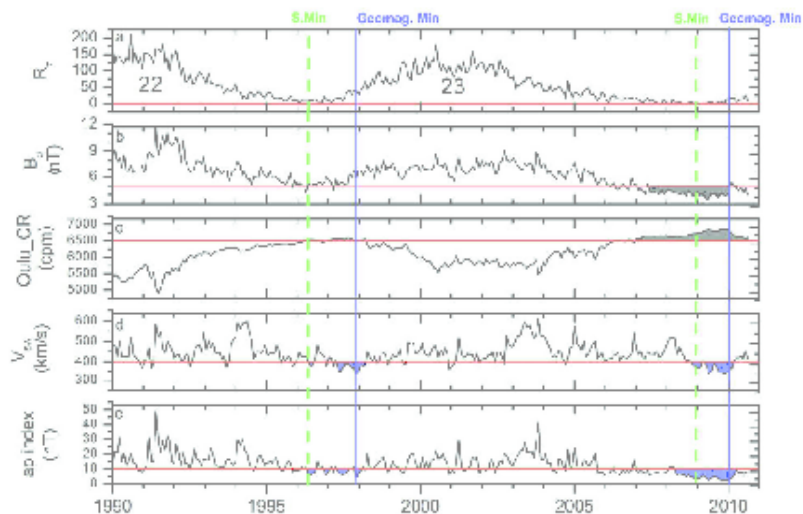

**Figure 3.** Interplanetary 1 AU near-Earth data for cycle 23 minimum (2008) compared to cycle 22 minimum (1996). The interplanetary magnetic field and solar wind speed are shown from 1990 through 2010 in the second and fourth panels from the top. The bottom panel give the geomagnetic ap index. The figure is taken from Tsurutani et al. (2011a).

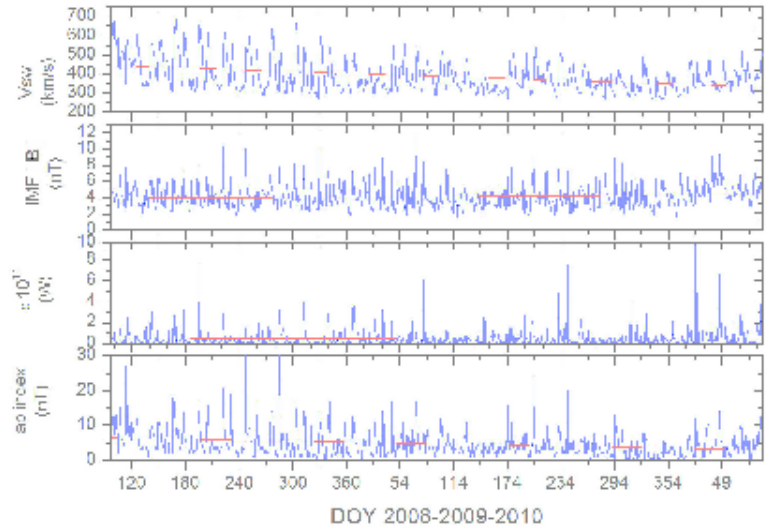

**Figure 4** A blow up of the 1 AU interplanetary parameters and the ap index.

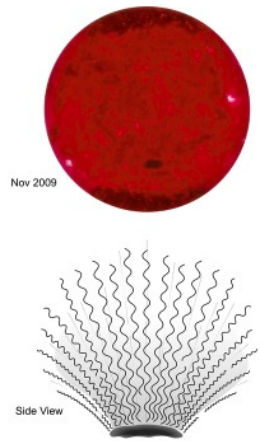

**Figure 5.** A coronal hole and Alven waves from it.   (top panel) A midlatitude coronal hole during Nov 2009. (bottom panel) a side view of the high speed solar wind coming from a coronal hole.   There is superradial expansion which leads to weaker speeds and smaller Alfvén wave amplitudes at the sides of the holes.

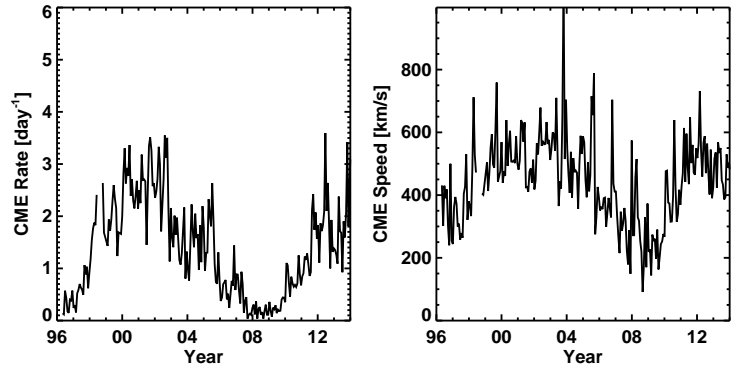

**Figure 6.** CME occurrence rate (per day) and speed in cycles 23 and 24 (1996 to 2013). The quantities have been averaged over Carrington rotation periods (27 days). Only CMEs with width ≥30° are included in the plot. The CME information is from the SOHO/LASCO CME catalog.

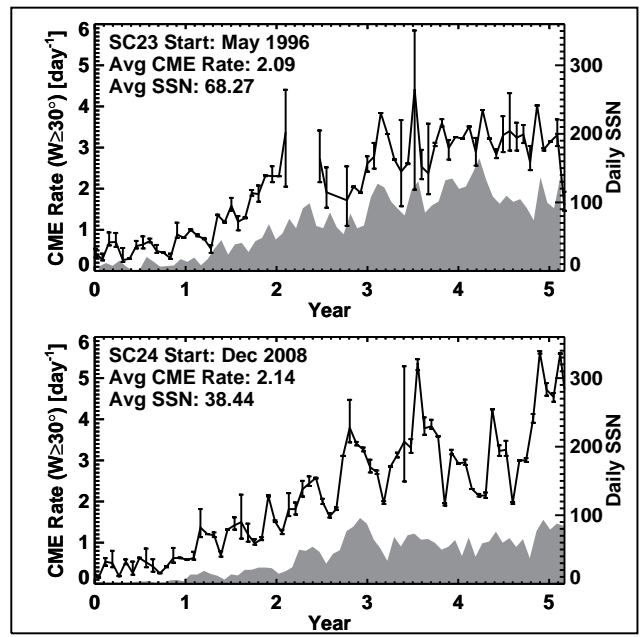

**Figure 7.** Detailed comparison between the corresponding epochs of cycles 23 and 24 (~5 years). Daily SSN (International) is included in each case. The average CME rates averaged over the first five years in each cycle are shown on the plots. The error bars are based on SOHO down times ≥3 h, which are listed in the SOHO/LASCO CME catalog.

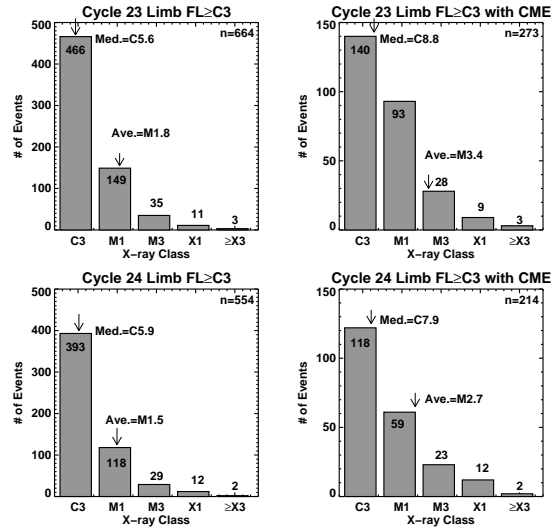

**Figure 8.** Distributions of flares originating within 30° from the limb for cycle 23 and 24. (left) all flares and (right) flares associated with CMEs. The average (Ave) and median (Med) values of the distributions are marked on the plots.

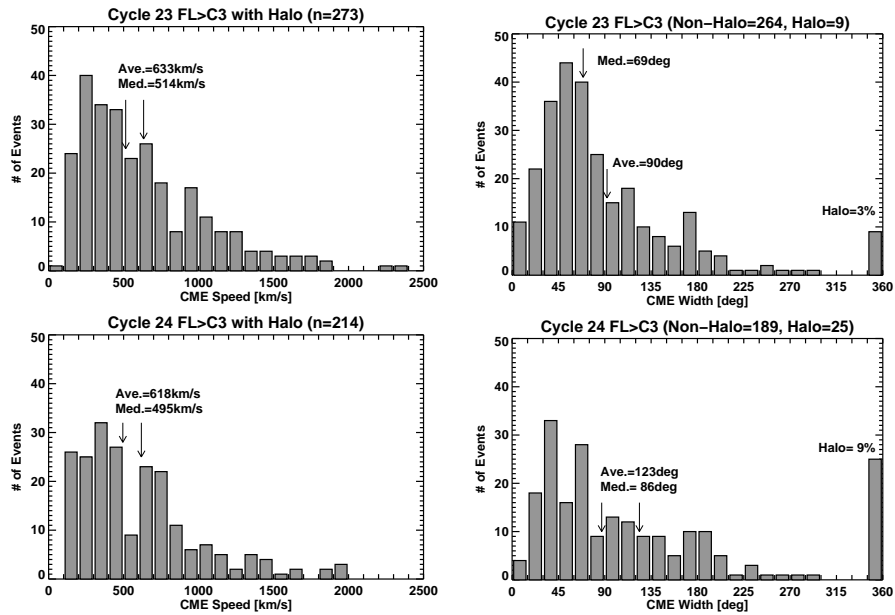

**Figure 9.** Speed and width distributions of the limb CMEs from cycle 23 (left) and 24 (right). The speed distributions are very similar, but the width distributions are different. The last bin in the width distributions represent full halo CMEs. Note that the full halo CMEs are three times more abundant in cycle 24.

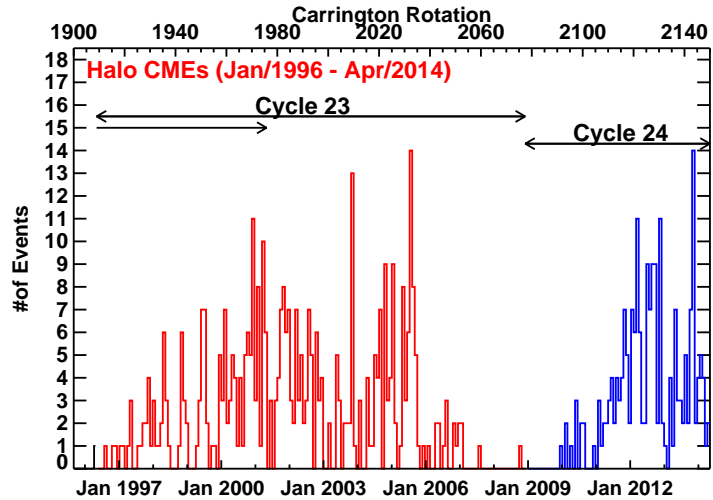

**Figure 10.** The occurrence rate of halo CMEs (number per Carrington rotation) from 1996 to the end of 2013 detected by SOHO/LASCO. From December 1, 2008 to December 31, 2013, there were 186 halo CMEs. Over the corresponding phase in cycle 23 (May 10, 1996 to June 9 2001), there were only 162 halos. If the halos occurred with the same average rate during the 4-month period when SOHO was operational, the number is expected to be ~173. Thus the number of full halos in cycle 24 is comparable to that of cycle23 or slightly greater (data from Gopalswamy et al. 2015).

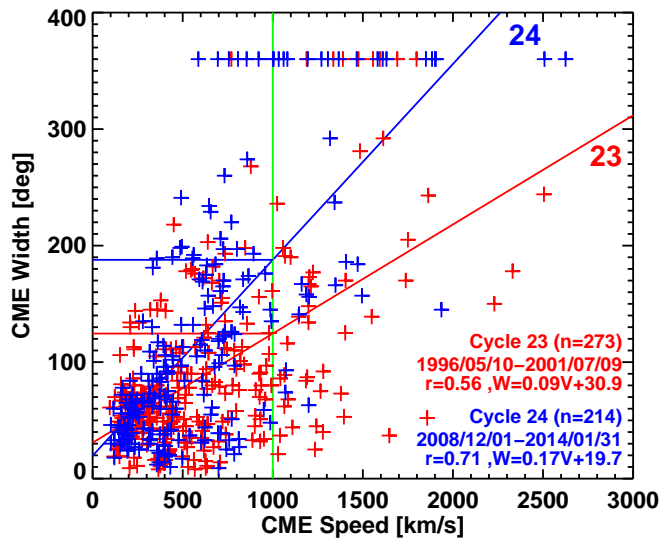

**Figure 11.** Speed vs. width distributions of limb CMEs from cycles 23 and 24. Both cycles show a good correlation between speed and width, but the slopes are very different. The correlations coefficients (r) and the regression lines are given on the plot. Student's t-test confirms that the slope difference is statistically significant. The data points at width=360° are halo CMEs, which are mostly from cycle 24.

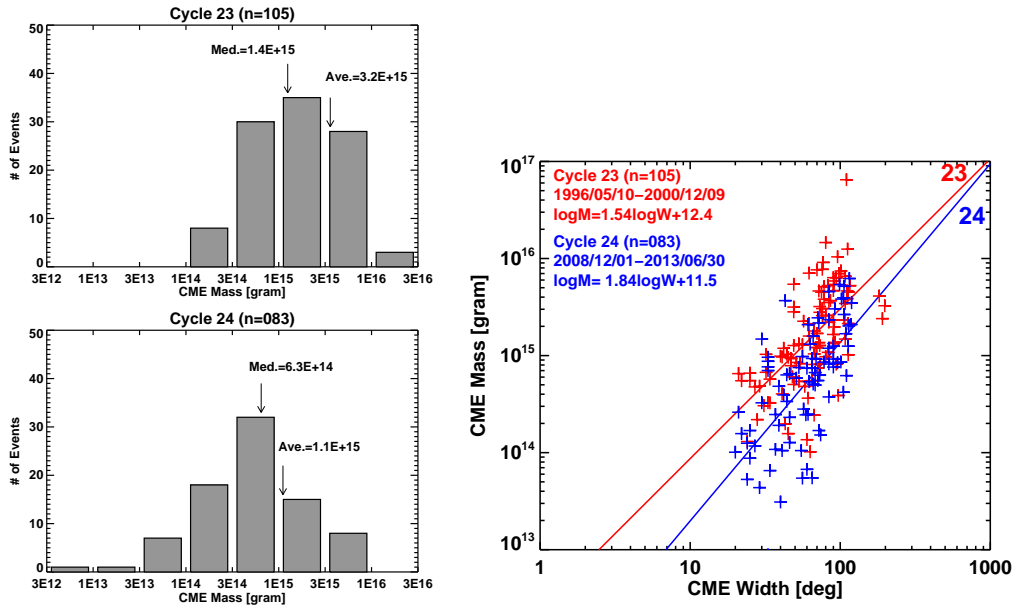

**Figure 12.** (left) CME mass distributions for limb CMEs from cycle 23 and 24. (right) Mass – width plots for the two cycles. For a given CME width, the mass is larger for cycle-23 CMEs. The mass estimate is restricted to the width range 20–120º (inclusive).

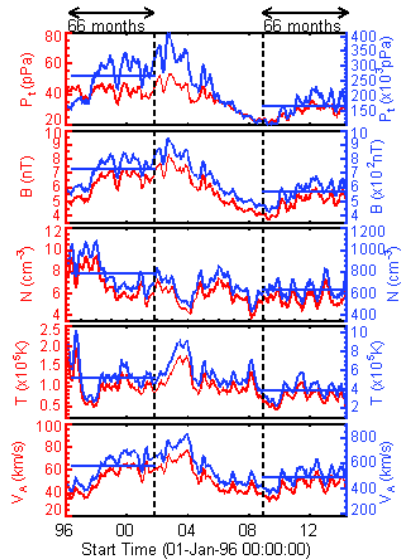

**Figure 13.** Physical parameters of the solar wind at 1 AU obtained from the OMNI data base from January 1996 through May 2014: Total pressure (Pt), magnetic field magnitude (B), proton density (N), proton temperature (T), and the Alfven speed (VA) at 1 AU (red lines with left-side Y-axis). The same quantities extrapolated from 1 AU to the corona (20 Rs) are shown by the blue lines (right-side Y-axis), assuming that B, N, and T vary with the heliocentric distance R as R-2, R-2, and R-0.7, respectively. The blue bars denote the 66-month averages in each panel, showing the decrease of all the parameters in cycle 24.

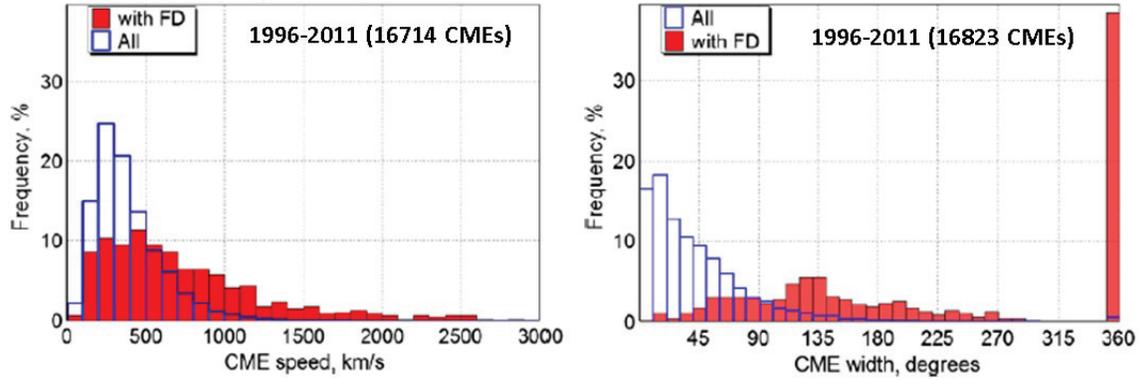

**Figure 14.** Comparison between the general population of CMEs (blue) and the FD-associated CMEs for CME speed (left) and width (right). All quantities were measured in the sky plane and no projection correction has been made (from Belov et al. 2014).

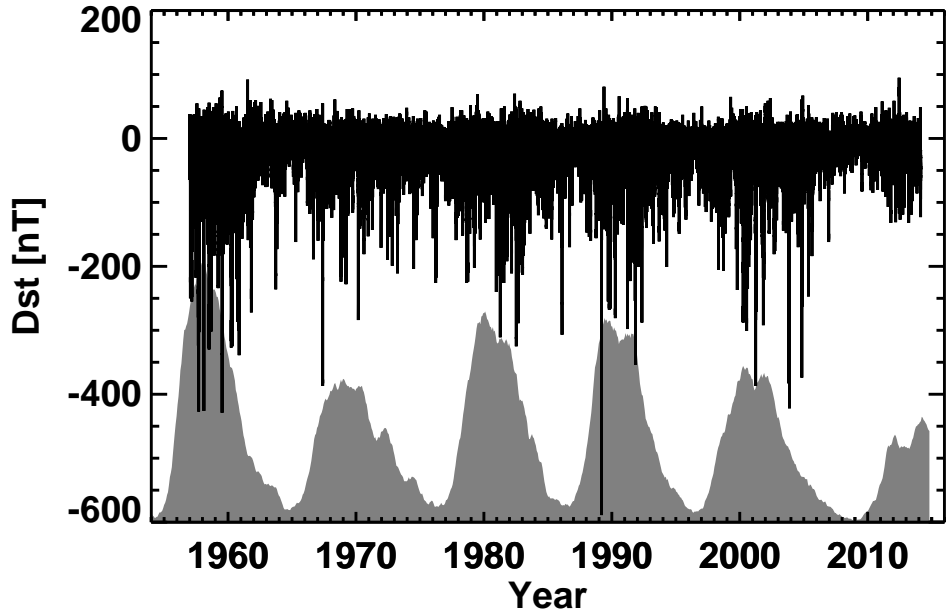

**Figure 15.** A plot of the Dst index since 1957 from the World Data Center, Kyoto (http://wdc.kugi.kyoto-u.ac.jp/dstdir/). The vertical streaks extending beyond -100 nT are the major storms. The largest storm (Dst ~ -589 nT) is the one on 1989 March 13 storm, which caused the power blackout in Quebec, Canada (Allen et al., 1989). The Halloween 2003 storm was the next largest storm (Dst < - 400 nT) after the Quebec storm (Mannucci et al., 2005). The positive excursions are the sudden commencements or sudden impulses caused by the impact of CME-driven shocks on the magnetosphere. The sunspot number is plotted at the bottom for reference.

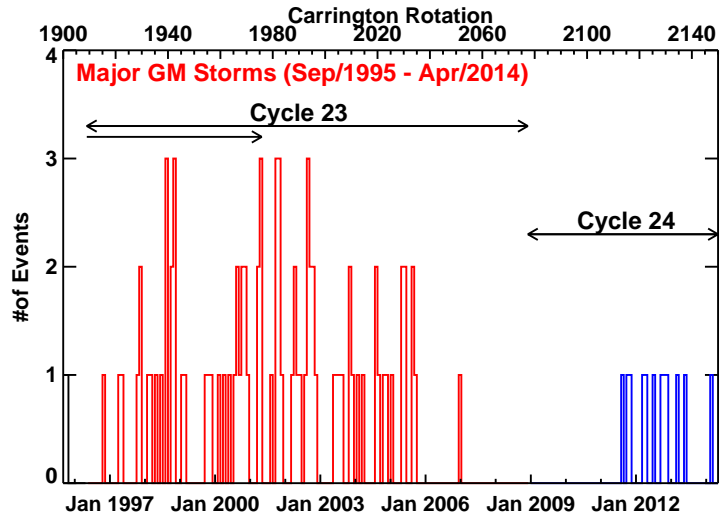

**Figure 16.** Number of major storms (Dst ≤ -100 nT) from Sep 1995 to April 2014. The black, red, and blue plots denote cycle 22, 23, and 24, respectively. The single arrow in the within the cycle-23 interval corresponds to the first 66 months of cycle 23 (May 1996 to November 2001) used for comparing with the current length of cycle 24 (December 2008 to April 2014). The first major storm occurred only in October 2011.

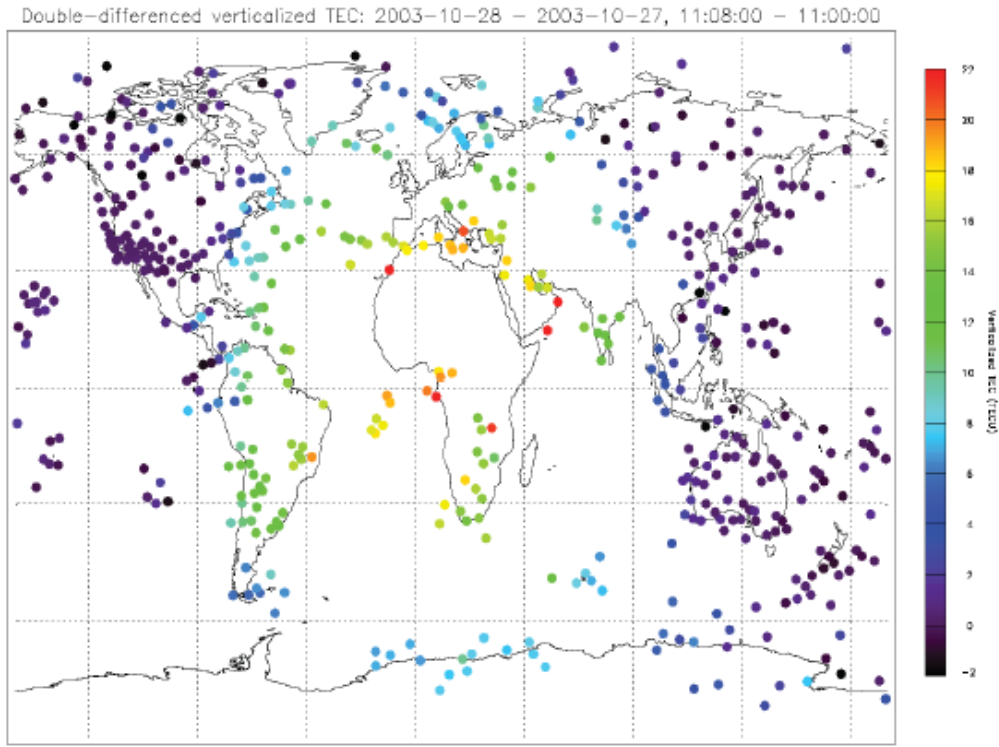

**Figure 17.** The ionospheric TEC change for the October 28, 2003 solar flare. The subsolar point is at the center of the graph. The figure is taken from Tsurutani et al. (2005).

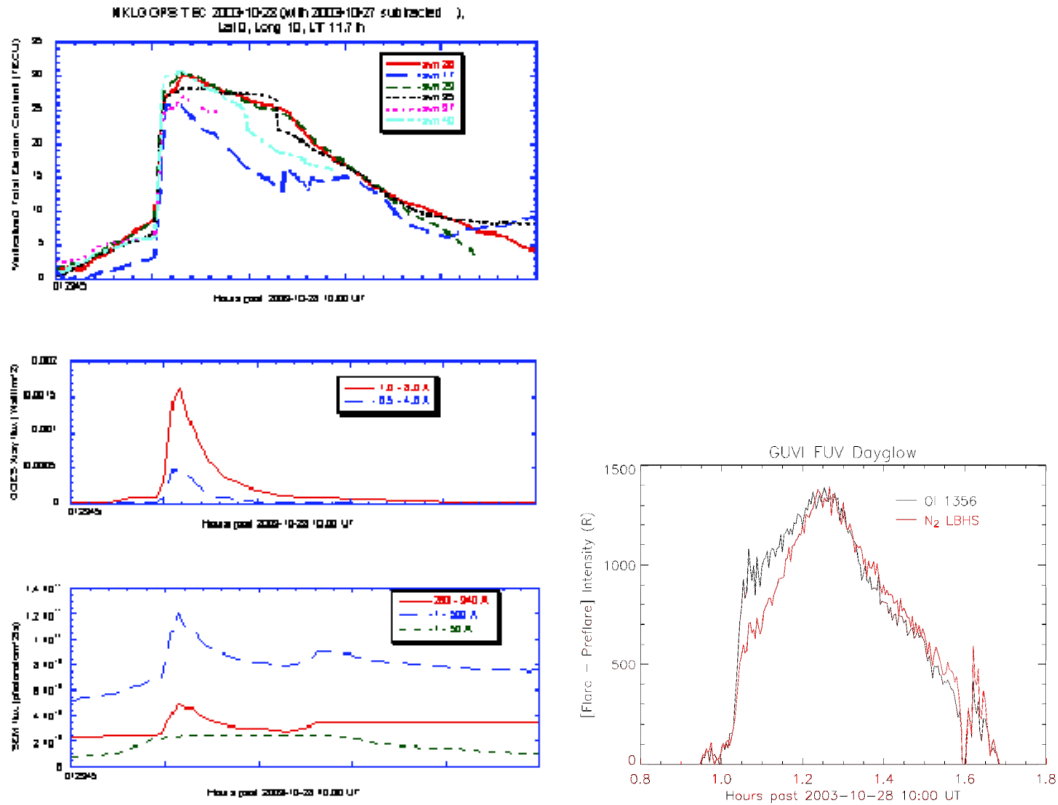

**Figure 18.** From top to bottom are the SOHO SEM FUV detector, a simulated X-ray flux profile, the Libreville Gabon (Africa) TEC determined from tracking 6 different GPS satellites and the O and N2 GUVI dayglow. This figure is taken from Tsurutani et al. (2005).

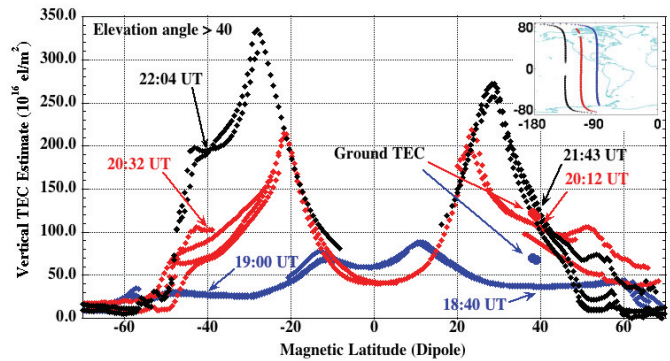

**Figure 19.** The vertical TEC above the CHAMP satellite (~400 km altitude) for the October 30, 2003 magnetic storm. The blue curve is the TEC prior to the magnetic storm. The red and black traces are the overhead TEC values after storm onset. (From Mannucci et al. 2005).

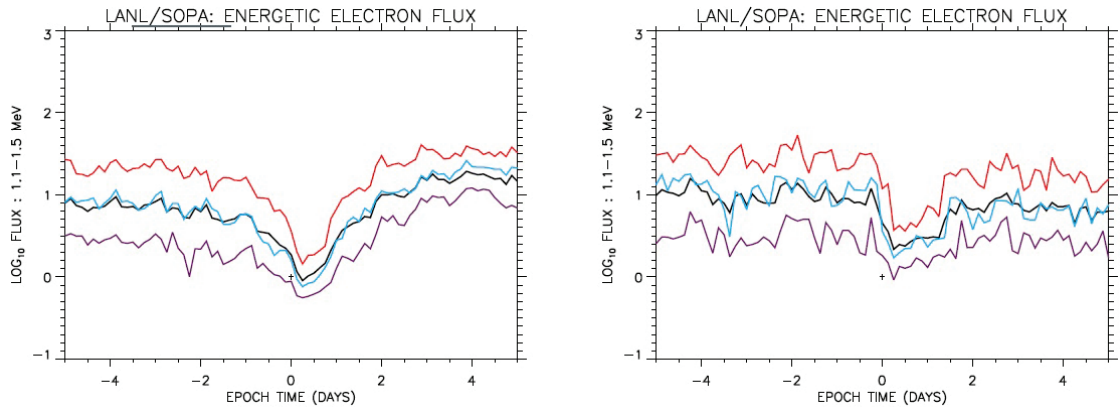

**Figure 20.** Energetic electron flux (1.1–1.5 MeV) at geosynchronous orbit for 93 strong HSSs (left) and 22 weak HSSs (right) presented from five days before to five days after zero epoch). The black and blue lines denote the mean and median values, respectively; the red and purple lines denote the upper and lower quartile, respectively. (from Denton and Borovsky 2012).

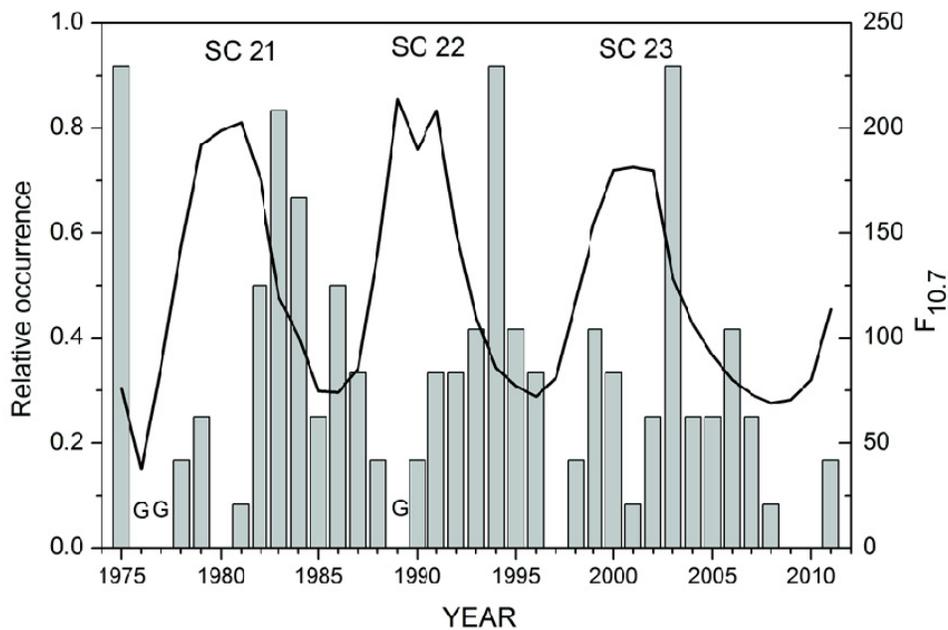

**Figure 21.** The solar cycle dependence of HILDCAAs. The relative occurrence of HILDCAAs during cycles 21, 22 and 23. The F10.7 solar flux is given as a black line. "G" represent data gaps. No AE data were available for the years 1976 and 1977. The figure is taken from Hajra et al. (2013).

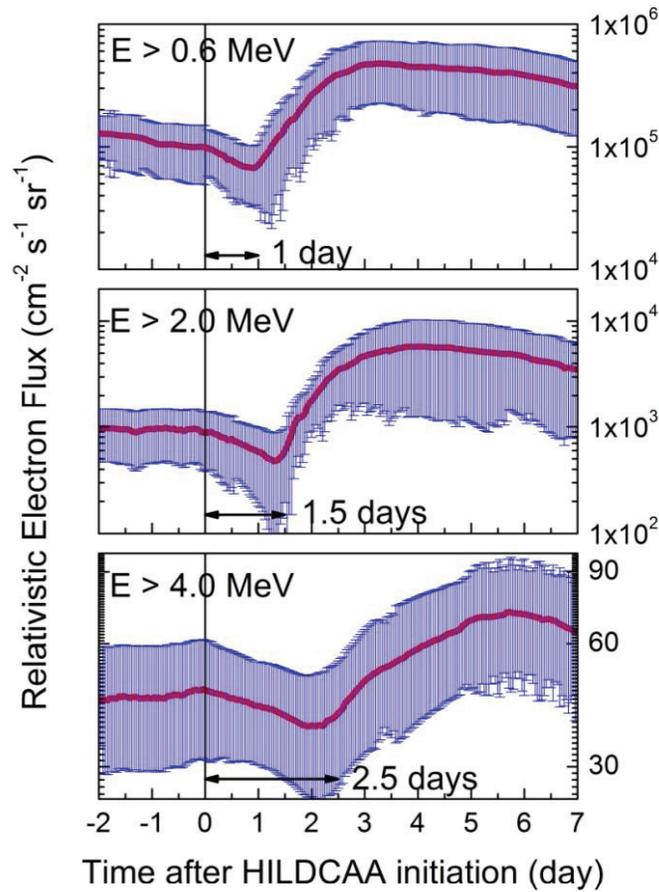

**Figure 22.** Relativistic electron fluxes ordered by HILDCAA onsets. The data used were all HILDAA events that occurred during cycle 23. The figure is taken from Hajra et al. (2014).

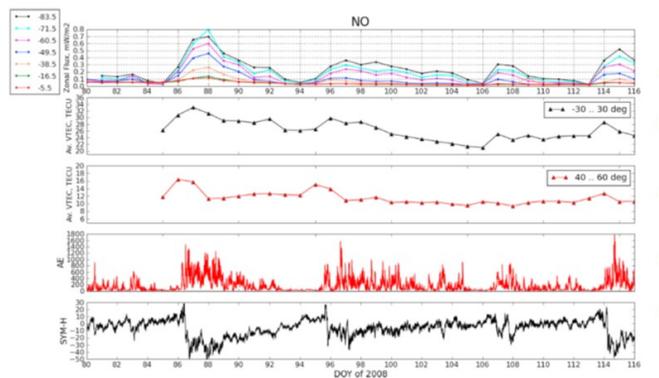

**Figure 23.** The WHI geoactive interval, 25 March to 26 April 2008. From top to bottom are the atmospheric NO IR irradiation over different latitude bins, the ionospheric vertical TEC near the equator, the equatorial VTEC at middle latitudes, the geomagnetic AE index and the SYM-H magnetic storm index. The figure is taken from Verkhoglyadova et al. (2011).

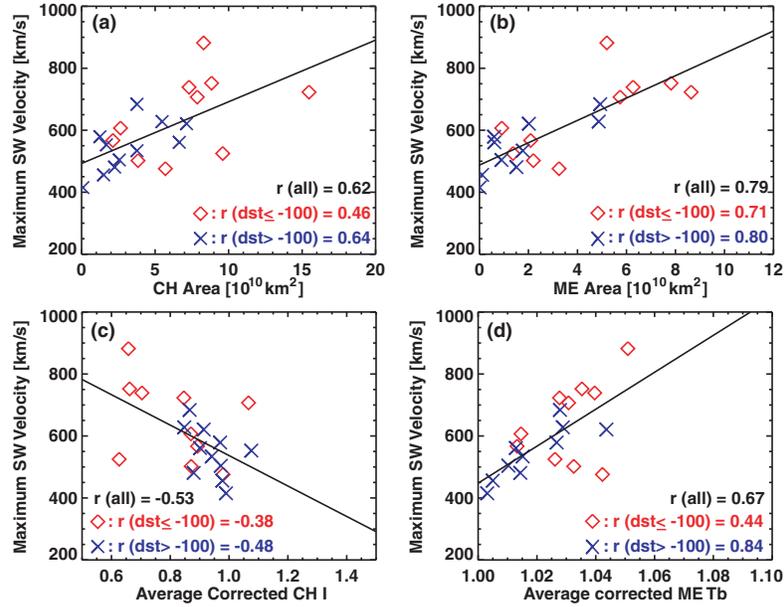

**Figure 24.** Scatter plots between solar wind (SW) speed and coronal hole parameters: (a) coronal hole area in EUV, (b) coronal hole area that shows microwave enhancement (ME), (c) EUV intensity averaged over the coronal hole area (c), and the microwave brightness temperature (Tb) averaged over the ME area (from Akiyama et al. 2013).

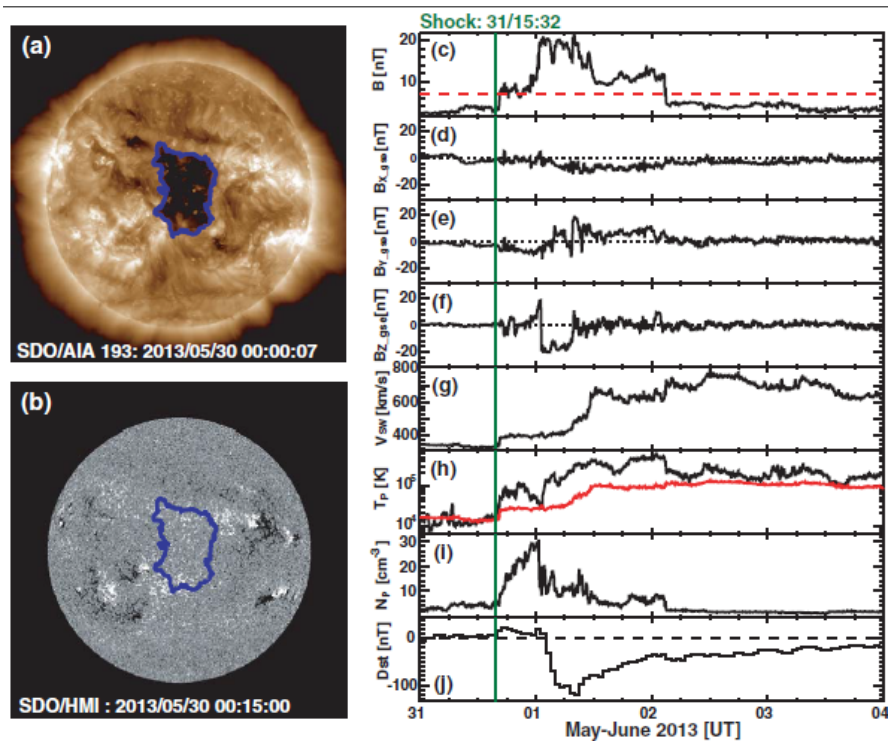

**Figure 25.** The first and only large CIR storm of cycle 24 as of this writing and its solar source region (coronal hole). The CIR was due to the high-speed stream from a large equatorial coronal hole (a) observed by SDO/AIA at 193 Å. At the photospheric level, the coronal hole had positive magnetic polarity (b) as observed by SDO/HMI. The outline of the coronal hole in (a) is overlaid on the magnetogram (b) to show the unipolar region. The CIR manifested as a region of enhanced magnetic field in in-situ observations (c). The three components of the IP magnetic field are shown in (d-e). The Bx component was negative indicating that the direction of the magnetic field was pointing away from the Sun consistent with the positive polarity in the photospheric magnetogram. The Bz component was negative for ~6h and was responsible for the major storm. It should be noted that this type of magnetic structure in CIRs is atypical (see previous discussion). The solar wind speed increased from ~380 km/s first as a shock jump and then at the interface and finally in the high speed stream (g). The temperature remained higher than the expected solar wind temperature (shown in red) throughout the interface and the fast wind (h). The density in the interface was high due to the shock compression reaching values in the range 10–30 cm-3. Finally the Dst index shows the evolution of the storm, reaching a peak value of -119 nT on June 1 at 9 UT (j).

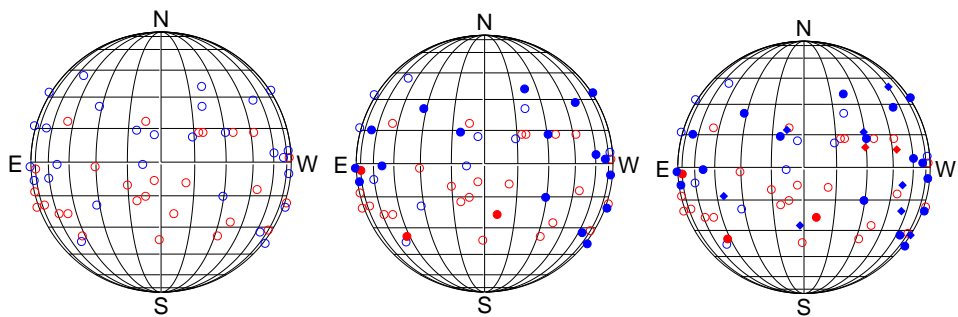

**Figure 26.** (left) solar sources of the 55 CMEs associated with major eruptions (flare size ≥M5.0) distinguished by their speed (blue ≥1500 km/s, and red <1500 km/s). (middle) filled circles denote the CMEs associated with large SEP events (3 red and 17 blue). (right) same as (b) but the sources of large SEP events associated with weaker eruptions (filled diamonds) are added, also divided into faster (blue, ≥1500 km/s) and slower (red, <1500 km/s). The speeds were obtained using a flux-rope fit to the CMEs using STEREO and SOHO data. The speed is not sky-plane projected.

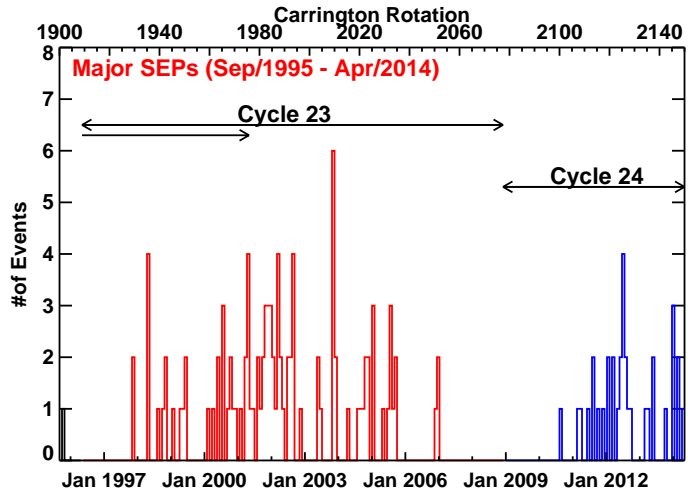

**Figure 27.** Number of large SEP events as a function of time from September 1995 to the end of April 2014. The black, red, and blue plots denote cycle 22, 23, and 24, respectively. The right-point arrow within the cycle-23 interval corresponds to the first 66 months of cycle 23 (May 1996 to November 2001) used for comparing with the current length of cycle 24 (December 2008 to April 2014). The first large SEP event occurred only in August 2010.

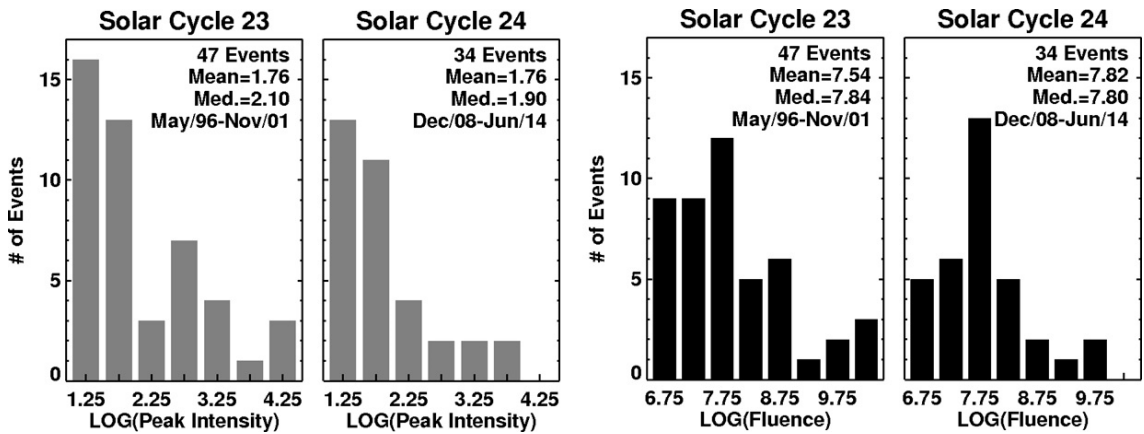

**Figure 28**. Peak-intensity (left) and fluence (right) distributions of large SEP events in the GOES >10 MeV channel compared between cycles 23 and 24. The mean and median values of the distributions did not differ significantly between the two cycles.

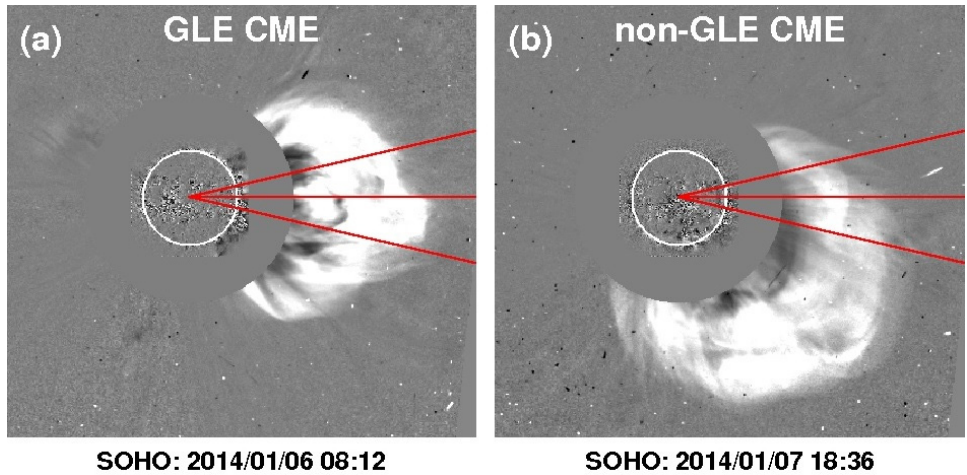

**Figure 29.** SOHO/LASCO snapshots of two CMEs, one associated with a GLE event (left) and the other with only a large SEP event. The January 6 CME occurred slightly behind the limb. The January 7 CME originated from close to the disk center. The three red lines mark the ecliptic (central line) and the edges of the ecliptic distance range (±13º) over which GLE CMEs occurred in cycle 23. The LASCO images are superposed on SDO/AIA difference images at 193 Å, which show the disturbances behind the west limb for the January 6 CME and close to the disk center for the January 7 CME. (adapted from Gopalswamy et al. 2014b).

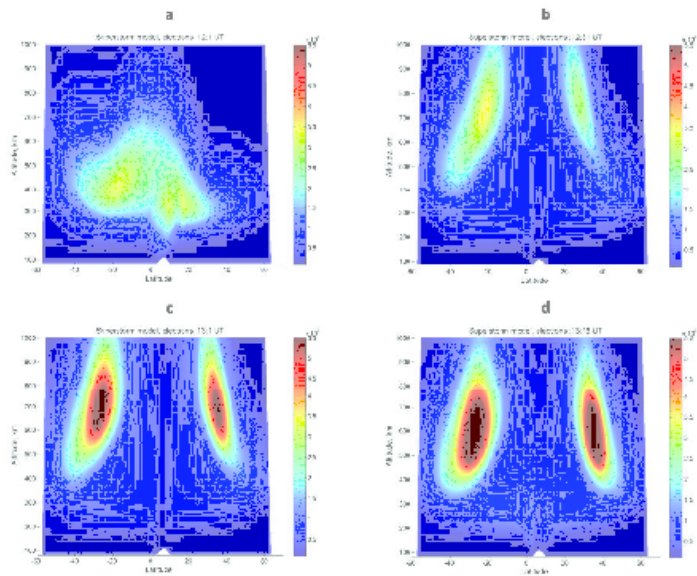

**Figure 30.** The oxygen ion profiles: a) prior to, b) 30 min after c) one hour after and d) 15 min after termination of the penetration of the interplanetary electric field. The figure is taken from Tsurutani et al. (2011).